\documentclass{aa}
\usepackage{graphicx}
\usepackage{subfigure}
\usepackage[]{natbib}
\bibpunct{(}{)}{;}{a}{}{,} 
\newcommand{\degree}{^{\circ}}
\newcommand{\MSun}{M_{\odot}}
\newcommand{\CGS}{\mathrm{erg \; cm^{-2} \; s^{-1}}}

\begin{document}
   \title{A year-long AGILE observation of Cygnus X-1 in hard spectral state}

   \author{E.~Del~Monte\inst{1}, M.~Feroci\inst{1}, Y.~Evangelista\inst{1,2}, E.~Costa\inst{1}, I.~Donnarumma\inst{1},
           I.~Lapshov\inst{1}, F.~Lazzarotto\inst{1}, L.~Pacciani\inst{1}, M.~Rapisarda\inst{3}, P.~Soffitta\inst{1},
           A.~Argan\inst{1}, G.~Barbiellini\inst{4,5}, F.~Boffelli\inst{6}, A.~Bulgarelli\inst{7}, P.~Caraveo\inst{8},
           P.W.~Cattaneo\inst{6}, A.~Chen\inst{8}, F.~D'Ammando\inst{1,10}, G.~Di~Cocco\inst{7}, F.~Fuschino\inst{7}, M.~Galli\inst{9},
           F.~Gianotti\inst{7}, A.~Giuliani\inst{8}, C.~Labanti\inst{7}, P.~Lipari\inst{2},
           F.~Longo\inst{4,5}, M.~Marisaldi\inst{7}, S.~Mereghetti\inst{8}, E.~Moretti\inst{4,5}, A.~Morselli\inst{10}, A.~Pellizzoni\inst{11},
           F.~Perotti\inst{8}, G.~Piano\inst{1,10}, P.~Picozza\inst{10}, M.~Pilia\inst{11,12}, M.~Prest\inst{12}, G.~Pucella\inst{3}, A.~Rappoldi\inst{6},
           S.~Sabatini\inst{1,10}, E.~Striani\inst{1}, M.~Tavani\inst{1,10}, M.~Trifoglio\inst{7}, A.~Trois\inst{1}, E.~Vallazza\inst{4},
           S.~Vercellone\inst{13}, V.~Vittorini\inst{1}, A.~Zambra\inst{14},
           L.~A.~Antonelli\inst{15,16}, S.~Cutini\inst{15,17}, C.~Pittori\inst{15,17}, B.~Preger\inst{15,17}, P.~Santolamazza\inst{15,17},
           F. Verrecchia\inst{15,17}, P.~Giommi\inst{15, 18}, L. Salotti\inst{18}}

   \offprints{E. Del Monte}

   \institute{INAF IASF Roma, Via Fosso del Cavaliere 100, I-00133 Roma, Italy\\ 
              \email{ettore.delmonte@iasf-roma.inaf.it}
              \and
              Dip. di Fisica, Universit\`a degli Studi di Roma ``La Sapienza'', P.le A. Moro 5, I-00185 Roma, Italy 
              \and
              ENEA, Via E. Fermi 45, I-00044 Frascati (Rm), Italy 
              \and
              INFN Trieste, Padriciano 99, I-34012 Trieste, Italy 
              \and
              Dip. di Fisica, Universit\`a di Trieste, Via Valerio 2, I-34127 Trieste, Italy
              \and
              INFN Pavia, Via Bassi, 6 I-27100 Pavia, Italy 
              \and
              INAF IASF Bologna, Via Gobetti 101, I-40129 Bologna, Italy 
              \and
              INAF IASF Milano, Via E. Bassini 15, I-20133 Milano, Italy 
              \and
              ENEA C.R. ``E. Clementel'', Via Martiri di Monte Sole 4, I-40129 Bologna, Italy 
              \and
              Dip. di Fisica, Universit\`a degli Studi di Roma ``Tor Vergata'',  Via della Ricerca Scientifica 1, I-00133 Roma, Italy 
              \and
              INAF Osservatorio Astronomico di Cagliari, loc. Poggio dei Pini, strada 54, I-09012, Capoterra (Ca), Italy 
              \and
              Dip. di Fisica e Matematica, Universit\`a dell'Insubria, Via Valleggio 11, I-20100 Como, Italy 
              \and
              INAF IASF Palermo, Via U.\ La Malfa 153, I-90146 Palermo, Italy 
              \and
              Consorzio Interuniversitario per la Fisica Spaziale, Viale Settimio Severo 63, I-10133 Torino, Italy 
              \and
              ASI Science Data Center, Via G.\ Galilei, I-00044 Frascati (Rm), Italy 
              \and
              INAF Osservatorio Astronomico di Roma, Via di Frascati 33, I-00040 Monte Porzio Catone (Rm), Italy 
              \and
              INAF staff resident at ASI Science Data Center 
              \and
              Agenzia Spaziale Italiana, Unit\`a Osservazione dell'Universo, Viale Liegi 26, 00198 Roma, Italy 
              }

\abstract
{Cygnus X-1 (Cyg X-1) is a high mass X-ray binary system, known to
be a black hole candidate and one of the brightest sources in the
X-ray sky, which shows both variability on all timescales and
frequent flares. The source spends most of the time in a hard
spectral state, dominated by a power-law emission, with occasional
transitions to the soft and intermediate states, where a strong
blackbody component emerges.}
{We present the observation of Cyg X-1 in a hard spectral state
performed during the AGILE science verification phase and
observing cycle 1 in hard X-rays (with SuperAGILE) and gamma rays
(with the gamma ray imaging detector) and lasting for about 160
days with a live time of $\sim 6$ Ms.}
{We investigated the variability of Cyg X-1 in hard X-rays on
different timescales, from $\sim 300$ s up to one day, and we
applied different tools of timing analysis, such as the
autocorrelation function, the first-order structure function, and
the Lomb-Scargle periodogram, to our data (from SuperAGILE) and to
the simultaneous data in soft X-rays (from RXTE/ASM). We concluded
our investigation with a search for emission in the energy range
above 100 MeV with the maximum likelihood technique.}
{In the hard X-ray band, the flux of Cyg X-1 shows its typical
erratic fluctuations on all timescales with variations of about a
factor of two that do not significantly affect the shape of the
energy spectrum. {From the first-order structure function, we find
that the X-ray emission of Cyg X-1 is characterized by
\textit{antipersistence} (anticorrelation in the time series, with
an increase in the emission likely followed by a decrease),
indicative of a negative feedback mechanism at work.} In the gamma
ray data a statistically significant point-like source at the
position of Cyg X-1 is not found, and the upper limit on the flux
is $\mathrm{5 \times 10^{-8} \; ph \; cm^{-2} \; s^{-1}}$ over the
whole observation (160 days). Finally we compared our upper limit
in gamma rays with the expectation of various models of the Cyg
X-1 emission, both of hadronic and leptonic origin, in the GeV --
TeV band.}
{The time history of Cyg X-1 in the hard X-ray band over 13 months
(not continuous) is shown. Different analysis tools do not provide
fully converging results of the characteristic timescales in the
system, suggesting that the timescales found in the structure
function are not intrinsic to the physics of the source. While Cyg
X-1 is not detected in gamma rays, our upper limit is a factor of
two lower than the EGRET one and is compatible with the
extrapolation of the flux measured by COMPTEL in the same spectral
state.}


\authorrunning {E. Del Monte et al.}
\titlerunning {The long AGILE observation of Cyg X-1}
\maketitle
%

\section{Introduction}

Cygnus X-1 (Cyg X-1) is one of the brightest X-ray sources in the
sky. It is a binary system composed of a compact object and the
O9.7 Iab supergiant star HDE 226868, filling $97 \; \%$ of its
Roche Lobe, with a mass ranging between $\sim 15$ and $\sim 30
\;\MSun$ \citep[see for
example][]{Gierlinski_et_al_1999,Caballero_et_al_2009}. The
measurement of the mass of the compact object, with a range
between 4.8 $\MSun$ and 14.7 $\MSun$ by \citet{Herrero_et_al_1995}
and $8.7 \pm 0.8 \; \MSun$ from
\citet{Shaposhnikov_Titarchuk_2007}, suggests identification with
a black hole. The distance to the binary system is measured as
$2.1 \pm 0.1$ kpc by \citet{Ziolkowski_2005}.

A characteristic feature of the black hole binaries (as Cyg X-1)
in the X-ray band, discussed for example by
\citet{Frontera_et_al_2001} between 0.5 and 200 keV, is the
existence of two well distinct emission states: ``low/hard'' and
``high/soft''. The typical energy spectrum in the low/hard state,
in which the source spends most of its time, is described well by
a power-law ($E^{-\Gamma}$) with photon index $\Gamma \sim$1.7 and
a high-energy cutoff at $\sim 150$ keV. Instead, in the high/soft
state the source is characterized by a strong blackbody component
with $kT \sim 0.5$ keV and a soft power-law tail with $\Gamma$
usually ranging between 2 and 3. An ``intermediate'' spectral
state also exists, discovered by \citet{Belloni_et_al_1996} in
observations with the Proportional Counter Array (PCA) aboard the
\textit{Rossi X-Ray Timing Explorer} (RXTE), in which the flux is
higher of about a factor of two with respect to the low/hard state
and the spectrum is softer (with a photon index of $\sim 2.1$ and
a blackbody component of $\sim 0.36$ keV temperature).

It is useful to note that the definition of low/hard and high/soft
states derives from the observations at soft X-rays. When
observing in hard X-rays, the condition reverses: the source is a
factor of $\sim$2 brighter in the low/hard state than in the
high/soft one. The typical flux in the low/hard state is $\sim 8
\times 10^{-9} \; \CGS$ in 20 -- 40 keV and $\sim 3 \times 10^{-8}
\; \CGS$ in 1 -- 10 keV. Although these definitions may be
misleading when applied to the observations in hard X-rays
reported in this paper, throughout the text we use the
classification low/hard versus high/soft to comply with the
classical literature on this source.

Cyg X-1 is a highly variable source in X-rays. Its variability is
observed on any timescale, from months to seconds \citep[see e.
g.][]{Brocksopp_et_al_1999,Pottschmidt_et_al_2003,Ling_et_al_1997}.
In particular, in the hard X-ray range the experiments of the
\textit{Interplanetary Network} detected seven episodes of giant
flaring in the 15 -- 300 keV energy range, with a duration of 0.9
to 28 ks, peak flux of order of $10^{-7} \; \mathrm{erg \; cm^{-2}
\; s^{-1}}$ and fluence ranging from $5 \times 10^{-5} \;
\mathrm{erg \; cm^{-2}}$ to $8 \times 10^{-4} \; \mathrm{erg \;
cm^{-2}}$ \citep[see][]{Golenetskii_et_al_2003}. These outbursts
were detected during both low/hard and high/soft spectral states,
and, in general, the giant bursting events seem to maintain the
spectral parameters (and likely the emission mechanisms) of the
underlying state.

Recently, Cyg X-1 has been observed by \citet{Albert_et_al_2007}
above 150 GeV energy with the \textit{Major Atmospheric Gamma
Imaging Cherenkov} (MAGIC) telescope. A TeV excess of 4.1 $\sigma$
compatible with a point-like source and spatially consistent with
Cyg X-1 was observed simultaneously with a hard X-ray flare taking
place in the low/hard state ($\sim$1.5 Crab - $1.2 \times 10^{-8}
\; \mathrm{erg \; cm^{-2} \; s^{-1}}$ - in 20 -- 40 keV with
INTEGRAL \citep{Malzac_et_al_2008} and
$\sim$1.8 Crab in 15 -- 50 keV of Swift/BAT and $\sim$0.6 Crab in
2 -- 12 keV of RXTE/ASM). The TeV excess was detected at the
rising edge of the hard X-ray peak, one day before its maximum,
while no variation is found in the soft X-ray emission. The source
does not show any steady emission in the TeV band and the upper
limits above $\sim 150$ GeV energy at the 95 \% confidence level
reach 2 \% of the Crab Nebula flux.

Although Cyg X-1 is a well-known source in X-rays, very little
information is known about its emission in gamma rays. During the
first three cycles of observation (1991 -- 1994) of the
\textit{Compton Gamma Ray Observatory} (CGRO), the source, in hard
state, was detected by COMPTEL only between 2 and 5 MeV
\citep[see][for details]{McConnell_et_al_2000}. EGRET did not
detect the source during that observation and the upper limit to
the flux is of the order of $10 \times 10^{-8} \; \mathrm{ph \;
cm^{-2} \; s^{-1}}$, posing no need for a high-energy cut-off. In
the soft state, the spectrum of Cyg X-1 can be modelled with a
power-law that extends with the same photon index ($\sim 2.5$ --
3) beyond 1 MeV and up to about 10 MeV as detected by COMPTEL
\citep{McConnell_et_al_2002}. Unfortunately in this case, no EGRET
measurement is available.

In steady conditions, the radio emission of Cyg X-1 is stable
during low/hard states \citep[see][]{Gleissner_et_al_2004}, except
for rarely observed flares \citep{Fender_et_al_2006}, and appears
to be quenched below a detectable level during the high/soft state
\citep{Brocksopp_et_al_1999}. A non-thermal radio jet, extending
up to $\sim15 \times 10 ^{-3}$ arcsec with an opening angle less
than 2$\degree$, was detected in VLBA observations by
\citet{Stirling_et_al_2001}.

The X-ray flares and a radio-emitting jet allow the classification
of Cyg X-1 as a microquasar, as described by
\citet{Mirabel_Rodriguez_1999}. The high-energy particles in the
radio-emitting jet are likely to produce gamma rays as well
\citep[see][]{Dubus_2007}. Following the model by
\citet{Zdziarski_Gierlinski_2004}, another possible source of
gamma rays, especially in the low/hard spectral state, is the
high-energy tail of the electron distribution in the corona.
Instead in the high/soft state, the energy spectrum does not show
an energy cutoff, as confirmed by the COMPTEL detection up to
$\sim 10$ MeV \citep{McConnell_et_al_2002} and a gamma ray
emission above this energy is expected.

AGILE \citep{Tavani_et_al_2008} is the first satellite mission
sensitive to gamma rays (in 30 MeV -- 30 GeV) flown after EGRET.
It observed Cyg X-1 for $\sim 160$ days in four different
pointings during its first observing cycle. Thanks to the X-ray
monitor SuperAGILE \citep{Feroci_et_al_2007}, the source is
observed at the same time in the hard X-ray band (between 20 and
50 keV). AGILE can continuously monitor the source, with a duty
cycle of about 50 \%. Other scanning instruments, such as RXTE/ASM
or Swift/BAT, observe each source many times a day but for a
shorter duration, and the duty cycle is usually shorter than 10
\%.

In this paper we report the probably longest uninterrupted
observation of Cyg X-1 in the hard X-ray and gamma ray bands from
the AGILE data, complemented and extended at lower energy (2 -- 12
keV) with the information from the public web
archive\footnote[1]{\texttt{http://heasarc.nasa.gov/xte\_weather/}}
\footnote[2]{\texttt{http://xte.mit.edu/ASM\_lc.html}} of the All
Sky Monitor (ASM) aboard RXTE. After the summary of the
observations (given in Sect. 2) and the description of the methods
of analysis (Sect. 3), we report the results of the data analysis
in X-rays (Sect. 4) and gamma rays (Sect. 5). Finally, in Sect. 6
we discuss our findings and draw our conclusions in Sect. 7.

\section{Observations}

AGILE observed the Cygnus region four times during the first cycle
of observations: from early November until mid December 2007 (44
days, hereafter observing block 1 or OB1), in the second and third
decades of April 2008 (20 days, hereafter observing block 2 or
OB2), in the second half of May and whole June 2008 (51 days,
hereafter observing block 3 or OB3), and from mid October until
the beginning of December 2008 (45 days, hereafter observing block
4 or OB4). The first observing block began during the satellite
science verification phase, which lasted from July until November
2007.

Each observation in the AGILE pointing plan usually lasts from 2
to 4 weeks to favour the photon collection in the gamma ray band.
The main constraint in the AGILE pointing strategy is the angle
between the Sun and the solar panels, which has to be fixed at $90
\degree \pm 1 \degree$. Because of this solar constraint, the
pointing of AGILE, hence the position of the sources in the field
of view, drifts by $\sim 1 \degree$ per day. For this reason the
off-axis angle (the distance between the source position and the
satellite boresight) is not constant during the observation. The
value of the off-axis angle of Cyg X-1 during the four observing
blocks ranges from $2.4 \degree$ to $32.7 \degree$. The exposure
of each observing block, as reported in the table, is computed as
the total time in which the source is observed, excluding the
duration of the satellite passages through the South Atlantic
Anomaly (SAA) and the source occultation by the Earth. Following
these criteria, the total on-source exposure amounts to $6.1
\times 10^6$~s. We show in table \ref{table:observation} the
details (start and end date, off-axis angle and exposure) of the
observing periods in the AGILE observation.

%
%
%
%


\begin{table*}[h!]
\caption{The observation log}             
\label{table:observation}      
\centering                          
\begin{tabular}{c c c c c}        
\hline\hline                 
Obs.   & Date & Modified   & off-axis angle & exposure \\    
period &      & Julian Day & [degrees]    & [ks]      \\    

\hline                        

OB1 & 02 Nov 2007 13:50 UT -- 16 Dec 2007 10:27 UT & 54406.576 -- 54450.435 & $ \;$ 2.4 -- 27.6 &      1634 \\
OB2 & 10 Apr 2008 15:10 UT -- 30 Apr 2008 11:15 UT & 54566.632 -- 54586.469 &      18.3 -- 15.4 & $ \;$ 893 \\
OB3 & 10 May 2008 17:52 UT -- 30 Jun 2008 11:10 UT & 54596.744 -- 54647.465 & $ \;$ 4.6 -- 32.7 &      1934 \\
OB4 & 17 Oct 2008 17:00 UT -- 01 Dec 2008 11:37 UT & 54756.708 -- 54801.484 &      15.1 -- 18.0 &      1632 \\


\hline                                   
\end{tabular}
\end{table*}

\section{Data reduction and analysis}

\subsection{X-rays}

SuperAGILE \citep[see][for a description]{Feroci_et_al_2007} is a
coded aperture instrument, and its data analysis software is
highly customized, as frequently happens for similar experiments.
The data of SuperAGILE are processed using the Enhanced Multi sky
Imaging (EMI) software package, suitable for decoding the images
of bright sources in moderately crowded fields \citep[see][for
details]{Feroci_et_al_2010}. In the observation reported in this
paper, Cyg X-1 is always the brightest source in the field of
view, making EMI a reliable method of data processing.

SuperAGILE is a 1-D imager with a field of view composed of two
orthogonal and overlapping regions, of $107 \degree \times 68
\degree$ each at zero response. In the central area, of $68
\degree \times 68 \degree$, both coordinates are encoded, thus
giving twice a 1-D imaging, and the sensitivity on axis is of the
order of 15 mCrab at $5 \sigma$ in one day with an energy
resolution of $\sim 8$ keV FWHM. During all the observations, Cyg
X-1 is in the twice 1-D region, with an off-axis angle ranging
from $2.4 \degree$ to $32.7 \degree$ (see table
\ref{table:observation}). As studied in the SuperAGILE ground and
in-flight calibrations
\citep[][]{Evangelista_et_al_2006,Feroci_et_al_2008}, the
variation of the instrument Point Spread Function in this range of
off-axis angles does not affect significantly the imaging
response.

The output of the EMI software, for all the sources detected in an
image above a significance level of $5 \sigma$, is the count rate
normalized with the instrument effective area convolved with a
Crab-like spectrum, making it independent of the source position
in the field of view. Throughout this paper we refer to the
normalized counting rate defined above, and as a reference, the
value for the Crab Nebula is $0.15 \; \mathrm{cts \; cm^{-2} \;
s^{-1}}$. The normalized rates from EMI have been calibrated by
means of a raster scan of pointings toward the Crab Nebula during
the AGILE science verification phase and also verified with other
bright sources using the Swift/BAT transient monitor
results\footnote[3]{\texttt{http://swift.gsfc.nasa.gov/docs/swift/results/transients/}}.
The uncertainty on the normalized rate values quoted throughout
this paper includes both the statistical and systematic components
\citep[see][for details]{Feroci_et_al_2010}. Since the systematic
uncertainty applies to the absolute value of the flux but not to
relative values of subsequent measurements in the same pointing,
in the plots we show only data points with statistical errors,
separately showing the systematic uncertainty on the flux
calibration.

We accumulated the ``correlated'' energy spectrum of Cyg X-1,
which is the flux of the source extracted from images integrated
in different energy intervals. The spectrum of Cyg X-1 was
extracted in bins of 2 keV amplitude, from 20 to 50 keV, by using
the EMI software. Since the source is bright, the significance of
the source in the images is greater than $5 \sigma$ in all the
energy intervals. The spectrum is accumulated separately in the
two 1-D directions and then fitted simultaneously using the
software package XSPEC v. 12.3.0 \citep{Dorman_et_al_2003}.

\subsection{Gamma rays}

We completed the study of Cyg X-1 with the observation by the
AGILE gamma ray imaging detector (GRID), sensitive in the energy
band between 30 MeV and 30 GeV. The data are analysed with the
AGILE Standard Analysis Pipeline \citep[see e. g.][for a detailed
description]{Vercellone_et_al_2008}. We accumulated the count map,
calculated the exposure map, and estimated the map of the gamma
rays from the Galactic Background only above 100 MeV and for
events flagged as confirmed gamma rays (\texttt{filtercode=5})
after excluding the data collected inside the SAA
(\texttt{phasecode=18}). The region of interest in the study of
Cyg X-1 reported in this paper has a $40\degree$ radius, inside a
map of $80\degree$ radius with $0.5 \degree$ bin size. We used a
particular application of the AGILE maximum likelihood (ALIKE)
procedure, called \texttt{ALIKEsingle}, to search for the
detection of sources at a specified position. Every run of the
\texttt{ALIKEsingle} software provides the user with the
significance of the maximum likelihood technique applied at the
input position and either the flux or the $2 \sigma$ upper limit
in case a point-like source is detected or not, respectively. To
sum all the data of the whole gamma ray observation, all the maps
are centred on the same galactic coordinates ($l=88.0 \degree$,
$b=-12.0 \degree$).

\section{Results in the X-ray band}

\subsection{The hard X-ray variability}

\begin{figure}[h!]
\centering
\includegraphics[angle=90, width=9. cm]{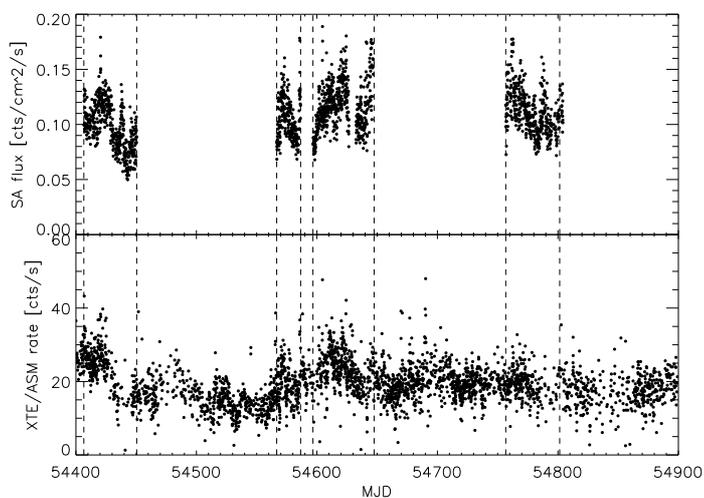}
\hfill \caption{Superposition of the SuperAGILE (upper panel) and
RXTE/ASM (lower panel) fluxes of Cyg X-1 during the observation.
The error bars are omitted for clarity and the RXTE/ASM data are
rebinned with 3 ks time bin. The vertical dashed lines limit the
observing blocks. The normalized rate of the Crab Nebula is $0.15
\; \mathrm{cts \; cm^{-2} \; s^{-1}}$ in SuperAGILE and 75
$\mathrm{cts \; s^{-1}}$ in RXTE/ASM.} \label{fig:SA-XTE}
\end{figure}

Thanks to the continuous monitoring of SuperAGILE, a measure of
the variability of Cyg X-1 with the resolution of about 3 ks from
the standard EMI pipeline is obtained. We adopted this resolution,
corresponding to an AGILE orbit, to guarantee the statistical
significance of the source detection in each time bin and to
favour the joint analysis with RXTE/ASM, although the SuperAGILE
data allow for finer time resolution. A synthetic view of the
normalized counting rate measured by SuperAGILE is given in Fig.
\ref{fig:SA-XTE}, together with the measure of RXTE/ASM in the
same time interval. The time bin of $\sim 3$ ks is used for both
instruments. As a reference, the counting rate of the Crab Nebula
is $0.15 \; \mathrm{cts \; cm^{-2} \; s^{-1}}$ in SuperAGILE and
75 $\mathrm{cts \; s^{-1}}$ in RXTE/ASM.

The continuous and uninterrupted source covering of SuperAGILE
allows us to detect possible flaring episodes and also to study
the flux with higher time resolution, if compared for example to
Swift/BAT with its higher sensitivity in a similar energy band (15
-- 50 keV) but a completely different pointing strategy with
shorter ``snapshots'' at higher sensitivity but lower duty cycle.

\begin{figure}[h!]
\centering
\includegraphics[angle=90, width=9. cm]{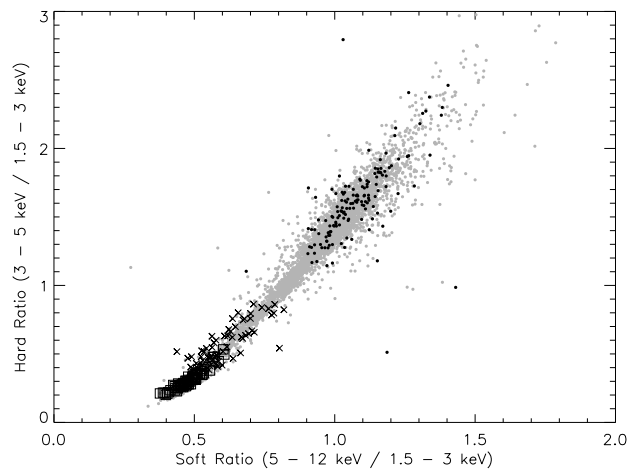}
\hfill \caption{Colour-colour diagram of Cyg X-1 from the public
data of RXTE/ASM with the hard ratio (estimated from the counting
rate in 5 -- 12 keV divided by the one in 1.5 -- 3 keV) as a
function of the soft ratio (from the fraction between the rate in
3 -- 5 keV and the one in 1.5 -- 3 keV), similarly to
\citet{Reig_et_al_2002}. The grey dots are the daily average of
the whole dataset (from MJD 50087 until 54801), the black dots are
the daily average of the data simultaneous to the AGILE
observation, the black squares are the data in the soft state from
MJD 50230 until 50307 \citep[see][]{Zdziarski_et_al_2002} and
finally the black crosses are the dwell averages in the
intermediate state from MJD 52797 until 52801
\citep[see][]{Malzac_et_al_2006} since this state lasted for only
five days.}. \label{fig:CCD}
\end{figure}

By comparing the rate of Cyg X-1 measured by RXTE/ASM during our
observation with typical values in the hard spectral state (20 --
30 $\mathrm{cts \; s^{-1}}$) and in the transitions to the soft
spectral state (more than 80 -- 100 $\mathrm{cts \; s^{-1}}$ on
June 1996 \citep[][]{Frontera_et_al_2001} or September 2001
\citep[][]{Pottschmidt_et_al_2003}), we derive that the source did
not reach the soft state during our observation. We study with
more completeness the spectral state during the AGILE observation
by means of a colour-colour diagram from the public data of
RXTE/ASM, plotted in Fig. \ref{fig:CCD}. In the figure the whole
dataset of RXTE/ASM (with daily integration between MJD 50087 and
54801) is represented by grey dots and we superimpose the data
simultaneous to the AGILE observation (black dots), the points
(squares) during the soft state between MJD 50230 and 50307
\citep[see][]{Zdziarski_et_al_2002} and during the intermediate
state of MJD 52797 -- 52801 (crosses), reported by
\citet{Malzac_et_al_2006}. From the distribution of the black
points in the colour-colour diagram, it is possible to see that
Cyg X-1 remained in the low/hard spectral state for the whole
duration of the AGILE observation.

\begin{figure*}[h!] 
\centering \subfigure{\includegraphics[angle=90,
width=.4\textwidth]{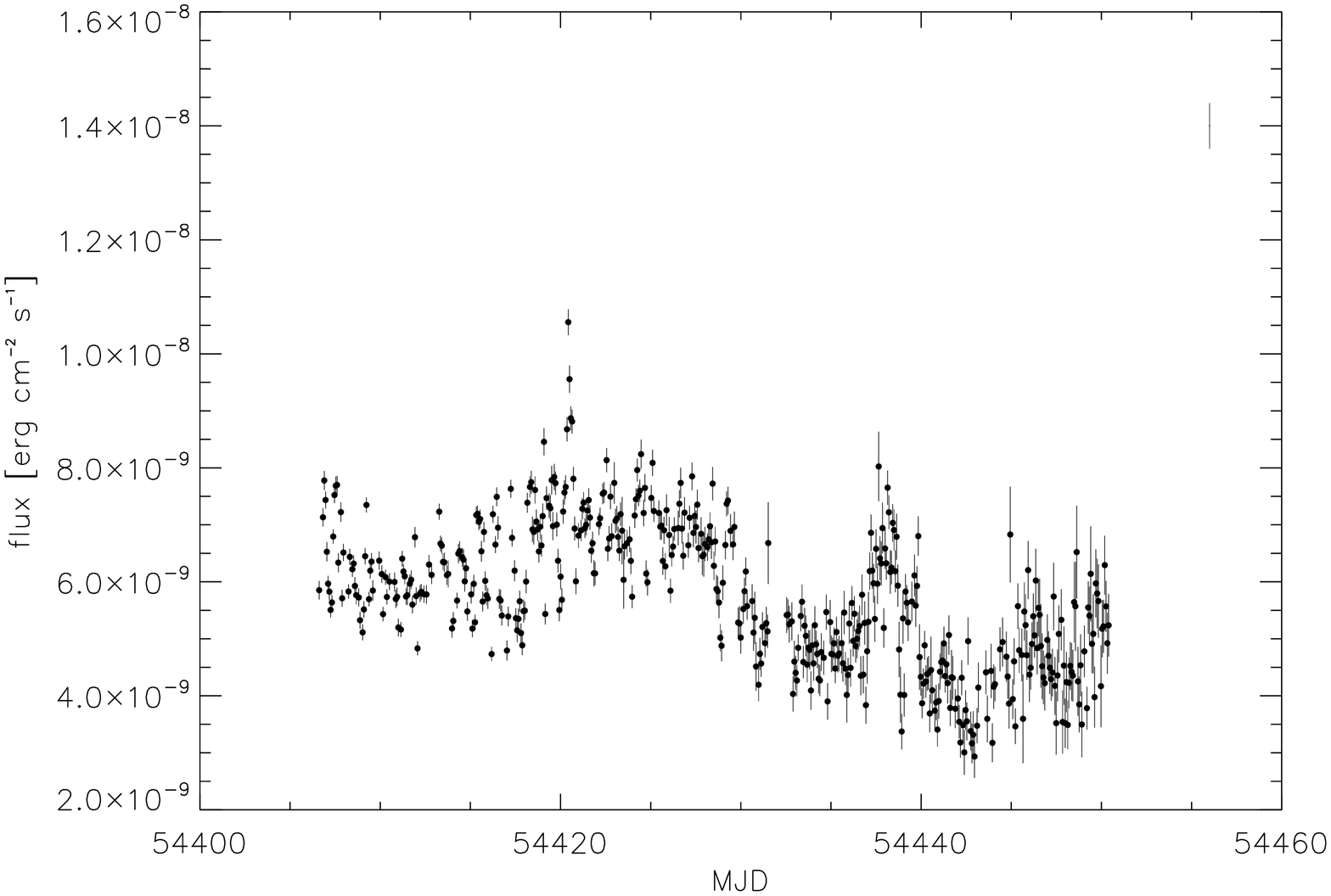}}
\subfigure{\includegraphics[angle=90,
width=.4\textwidth]{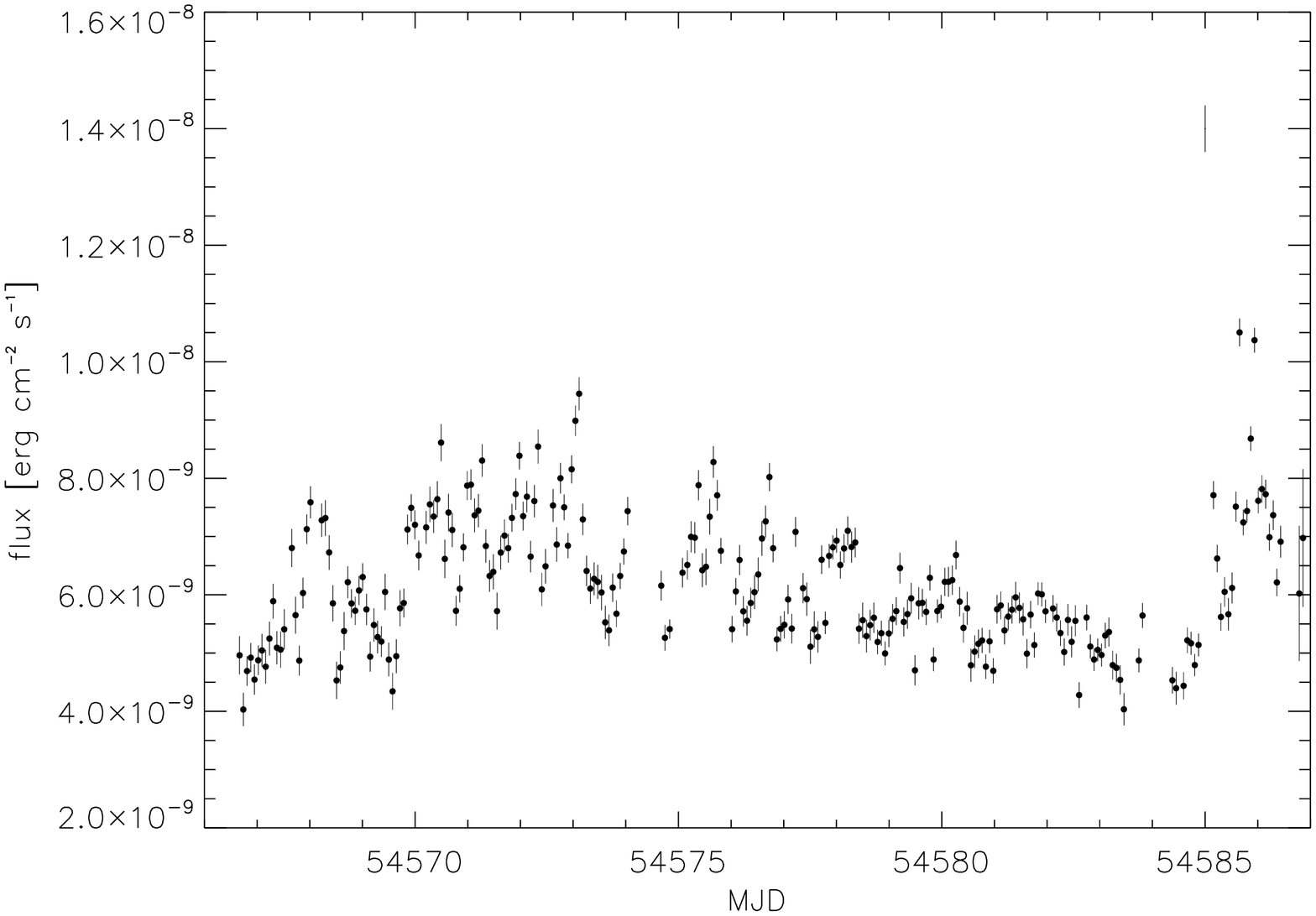}} \subfigure{
\includegraphics[angle=90, width=.4\textwidth]{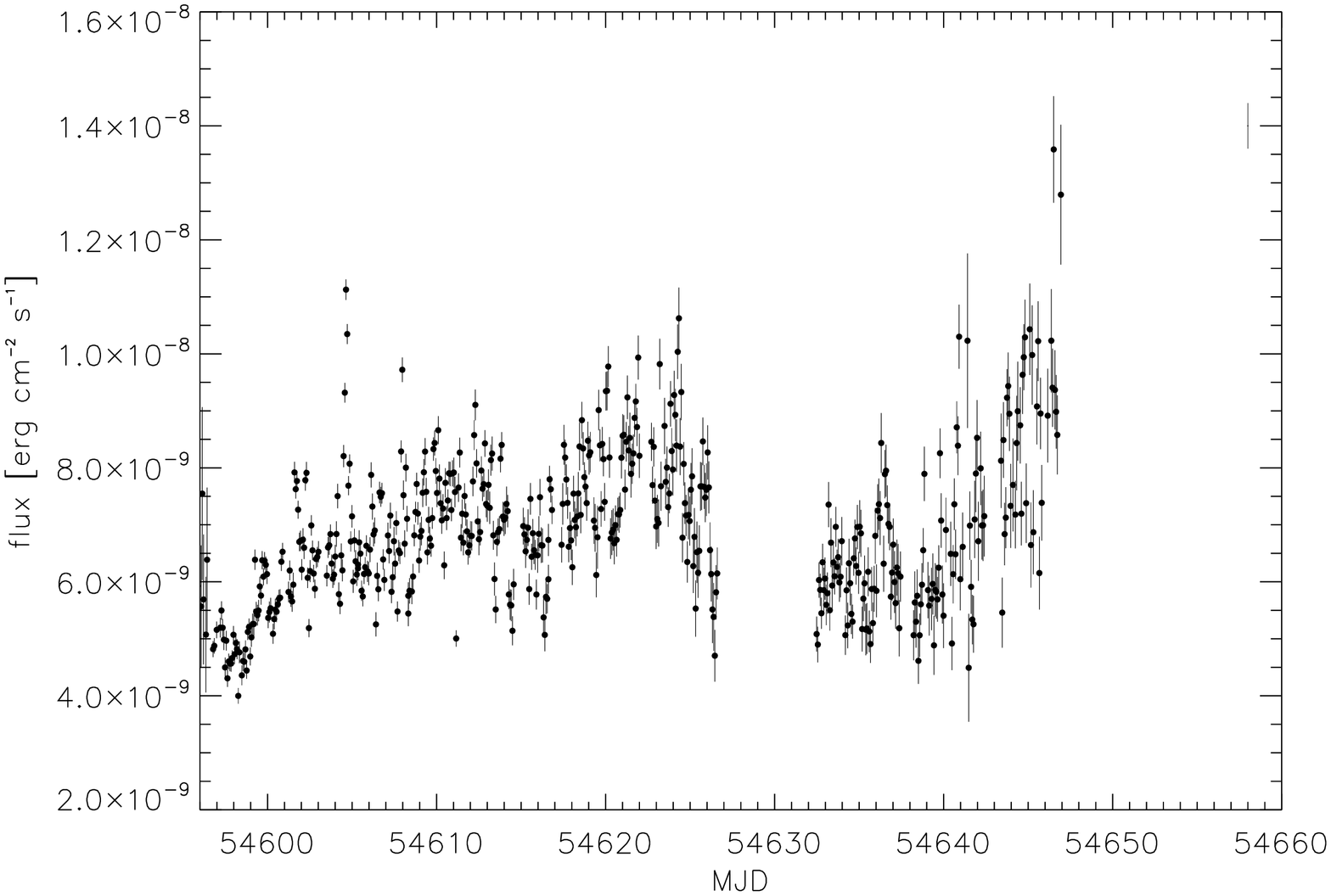}} \subfigure{
\includegraphics[angle=90, width=.4\textwidth]{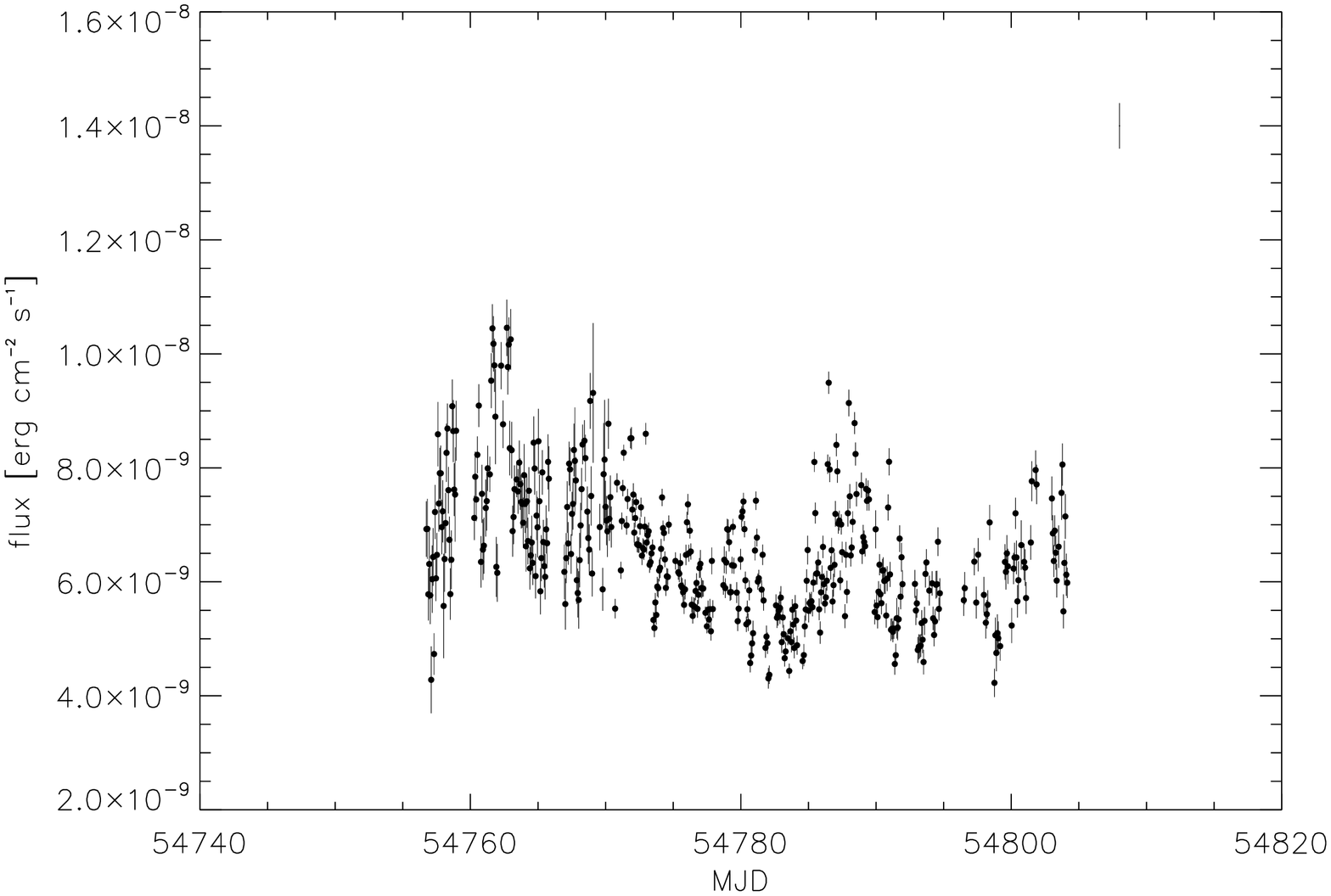}} \caption{SuperAGILE lightcurve of
Cyg X-1 during the observing periods 1 (top left), 2 (top right),
3 (bottom left) and 4 (bottom right). Each point represents the
normalized rate estimated in one satellite orbit (with exposure of
$\sim 3$ ks). The normalized rate values are converted into units
of $\CGS$ considering the source average spectrum. Only the
statistical uncertainty is shown on the points in the plots. The
average systematic uncertainty on the flux calibration is plotted
in the top right angle of each plot.}
\label{fig:SuperAGILE_orbital}
\end{figure*}

We select two different timescales to estimate the source counting
rate in the SuperAGILE band: one satellite orbit, with net
exposure of $\sim 3$ ks, and one day, with exposure of $\sim 40$
ks. The normalized rate of Cyg X-1 on orbital timescale is plotted
in Fig. \ref{fig:SuperAGILE_orbital}. All the plots have the same
vertical scale to compare the source variability. The rate is
converted into units of $\CGS$ using a conversion factor derived
from the fit of the average energy spectrum.

In the plots the characteristic slow variations of the emission of
Cyg X-1 are clearly detected, with a typical timescale of 30 -- 50
days in which the source flux changes up to a factor of $\sim 3$.
Superimposed to this slow modulation, a faster and erratic
variability, with timescale of about one day and a maximum flux
variation of the order of two, can be seen. In the same figures it
is also possible to appreciate the short time flickering of Cyg
X-1, on the scale of a few hours. Although some periods of rapid
flux increase were detected, for example around MJD 54420, 54605
or 54648, large flares were not detected during the AGILE
observation.

\begin{figure}[h!]
\centering
\includegraphics[angle=90, width=9. cm]{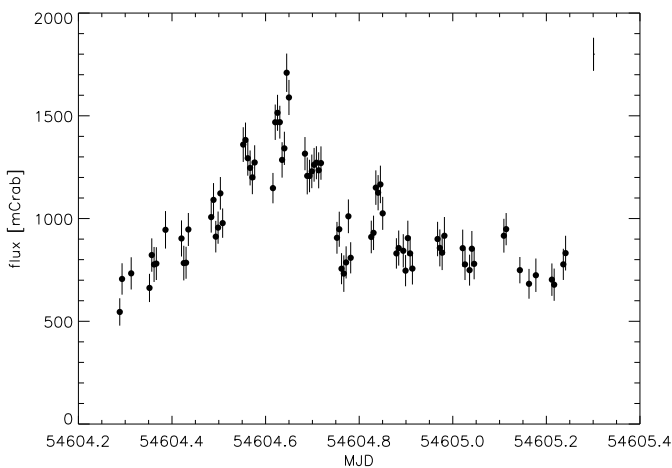}
\caption{SuperAGILE lightcurve of the source flux increase between
MJD 54604 and 54605 in the energy interval 17 -- 50 keV with a bin
size of $\sim 300$ s. The gaps in the lightcurve are due to the
passage of the satellite through the SAA and to the source
occultation by the Earth. Only the statistical uncertainty is
shown on the points in the plots. The average systematic
uncertainty on the flux calibration is plotted in the top right
angle.}\label{fig:SuperAGILE_hires_1}
\end{figure}

SuperAGILE always stores and transmits all the detected counts
(photon-by-photon operative mode). We could then measure the
normalized rate of Cyg X-1 accumulating images with a time bin of
$\sim 300$ s exposure, a trade-off between time and statistical
quality of the images. We excluded the time intervals
corresponding to the SAA and to Earth occultation. Usually we cut
the SuperAGILE data with an energy threshold fixed at 20 keV to
avoid the systematics due to the variation of the instrument
temperature. Instead, in the integration time of $\sim 300$ s the
temperature, and consequently threshold, variation is small, thus
we can lower the energy threshold down to 17 keV. We show for
example in Fig. \ref{fig:SuperAGILE_hires_1} the lightcurve  of
Cyg X-1 with a resolution of $\sim 300$ s between MJD 54604 and
54605, during a flaring episode. The higher time resolution allows
the short-time variability superimposed to the main flux variation
to be appreciated.

\begin{figure}[h!]
\centering
\includegraphics[angle=-90, width=9. cm]{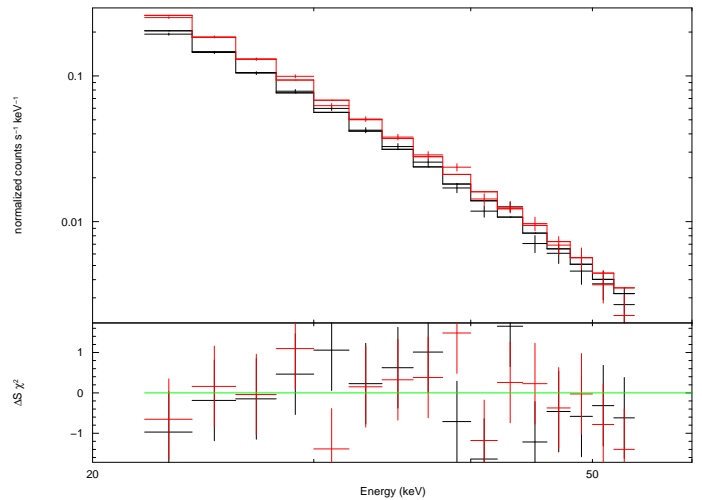}
\caption{Spectrum of Cyg X-1 from the SuperAGILE images in the X
direction showing the bright (MJD 54761 -- 54762, 75 ks exposure,
$0.17 \; \mathrm{cts \; cm^{-2} \; s^{-1}}$, red in colour
version) and faint (MJD 54597 -- 54599, 92 ks exposure, $0.08 \;
\mathrm{cts \; cm^{-2} \; s^{-1}}$, black in colour version) flux
states of the source. The parameters of the fit for the images in
both directions are reported in Table \ref{table:spectra}, where
the uncertainties are at a 90 \% confidence level and include a 5
\% systematic error.} \label{fig:spectrum_average}
\end{figure}

The availability of the photon-by-photon data also permits
building of the energy spectra. To check for spectral variability
possibly associated with the observed flux variations, we studied
the energy spectrum of Cyg X-1 in three flux states: average (from
MJD 54610 until 54612), faint (between MJD 54597 and 54599), and
bright (MJD from 54761 to 54762). A simple power-law, the model
\texttt{pegpwrlw} of XSPEC v.12.3.0 \citep{Dorman_et_al_2003},
provides a good fit to the spectrum in the 20 -- 50 keV energy
band (see Fig. \ref{fig:spectrum_average}). The results of the
spectral fit are summarized in Table \ref{table:spectra}. The
spectral shape of the source does not show significant variations
despite a flux variation of nearly a factor of two, except for a
marginal indication of a small softening at lower flux.

%
\begin{table*}[p]
\caption{Parameters of the spectral fit of Cyg X-1. $\chi^2_r$ is
the reduced chi square, the uncertainties are at 90 \% confidence level and include a 5 \% systematic error.}             
\label{table:spectra}      
\centering                          
\begin{tabular}{c c c c c}        
\hline\hline                 
Normalized rate & photon index & flux & $\chi^2_r$ & degrees of freedom \\    
$[\mathrm{cts \; cm^{-2} \; s^{-1}}]$ & & $[\mathrm{erg \; cm^{-2} \; s^{-1}}]$ & &\\%
\hline                        

$0.12$ & $1.63^{+0.09}_{-0.08}$ & $(7.6 \pm 0.3) \times 10^{-9}$ & 1.159 & 31 \\
$0.17$ & $1.60 \pm 0.11$ & $(8.3 \pm 0.4) \times 10^{-9}$ & 1.113 & 32\\
$0.08$ & $1.74 \pm 0.09$ & $(4.6 \pm 0.4) \times 10^{-9}$ & 0.986 & 29 \\

\hline                                   
\end{tabular}
\end{table*}

\subsection{Correlation between soft and hard X-rays}

\begin{figure*}[h!]
\centering \subfigure{\includegraphics[angle=90,
width=.4\textwidth]{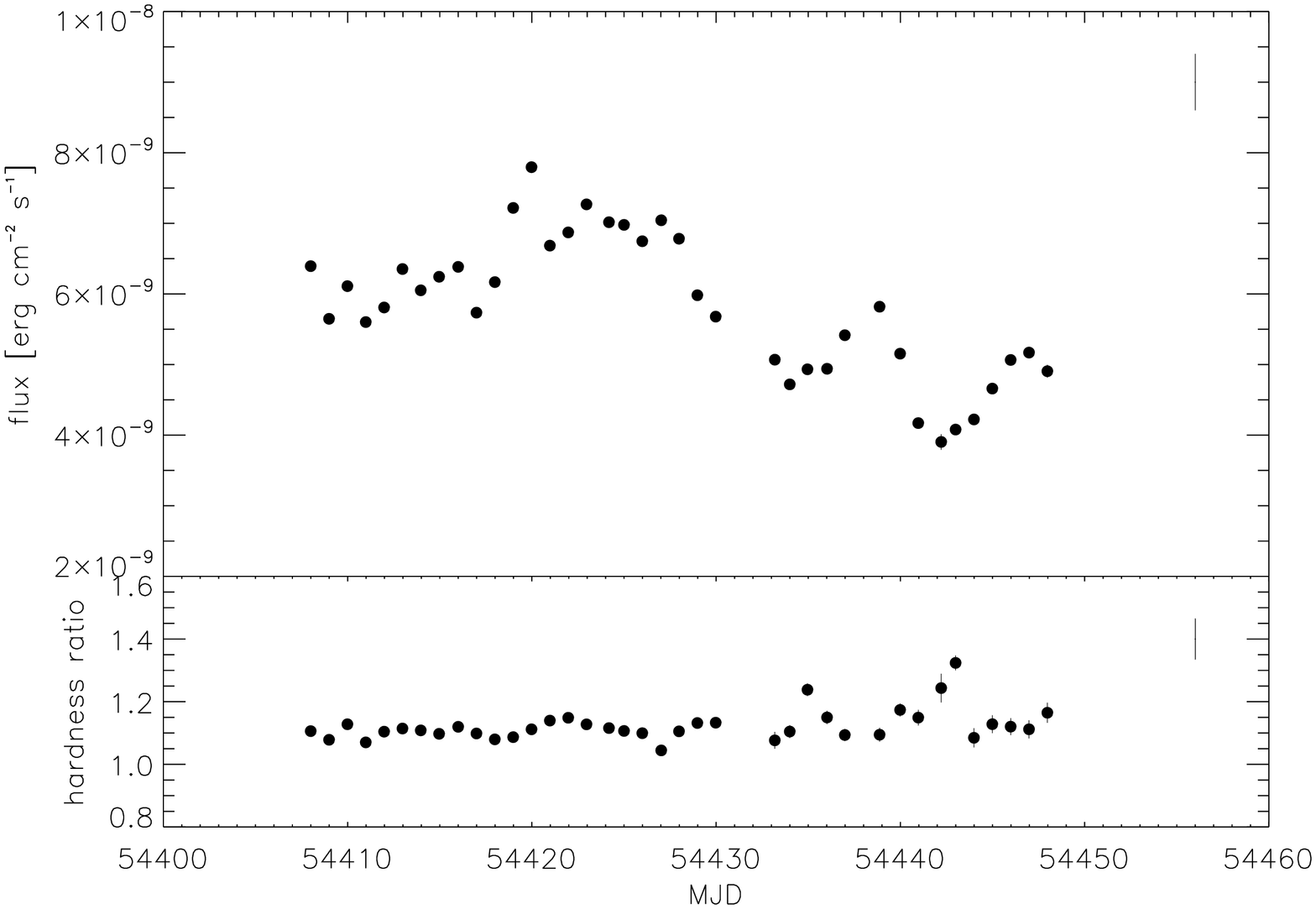}} \centering
\subfigure{\includegraphics[angle=90,
width=.4\textwidth]{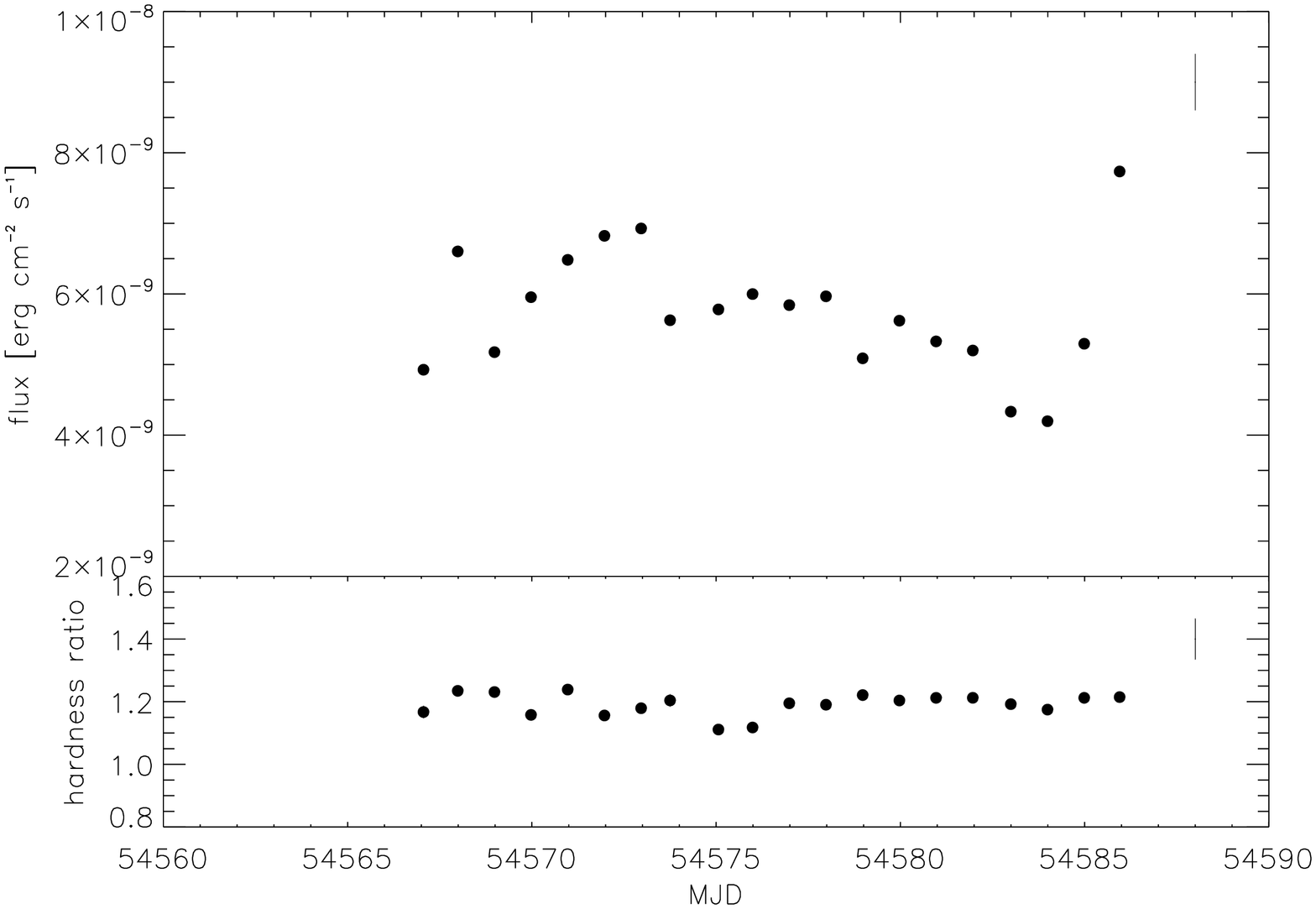}} \centering
\subfigure{\includegraphics[angle=90, width=.4\textwidth
]{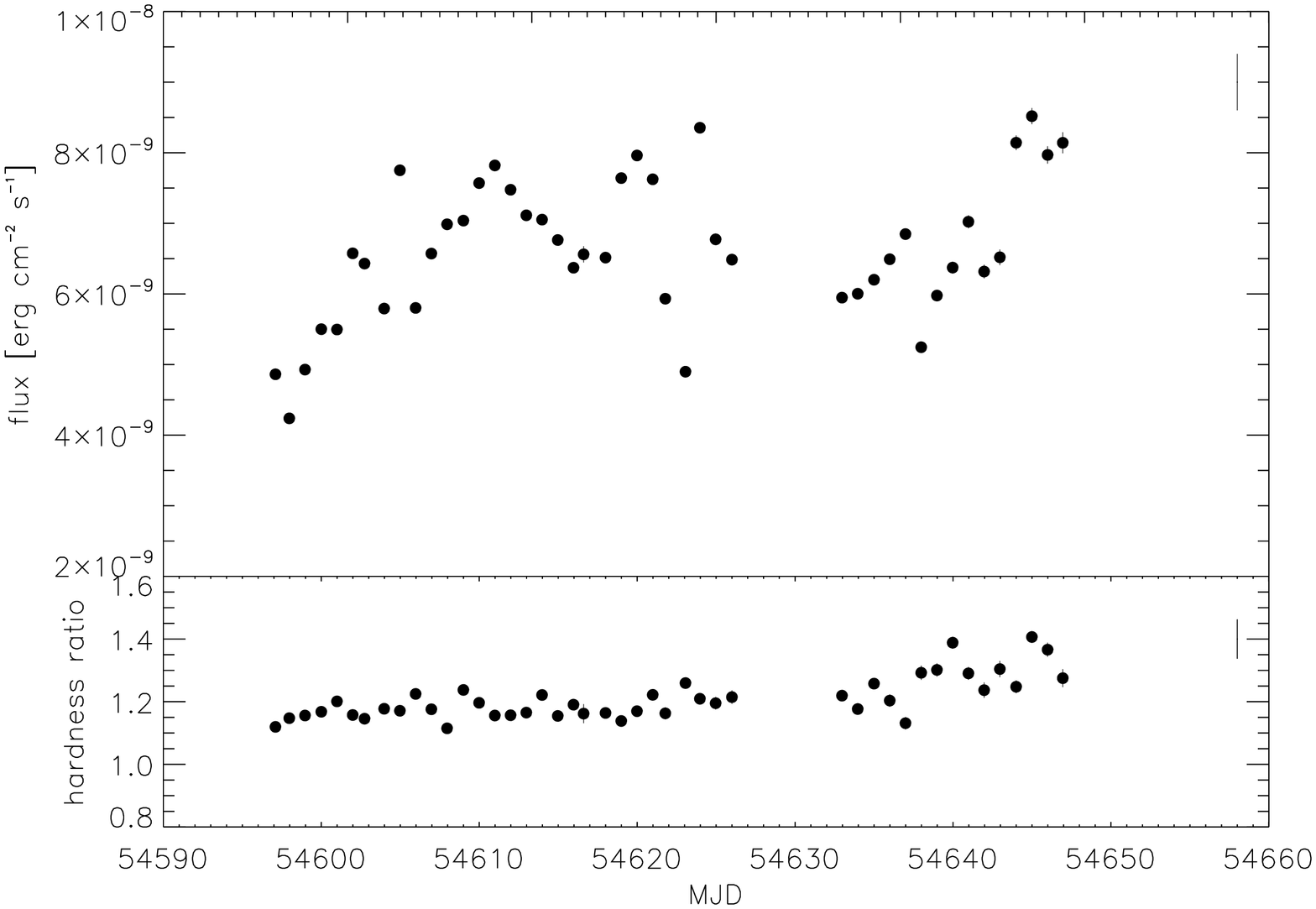}} \centering \subfigure{\includegraphics[angle=90,
width=.4\textwidth ]{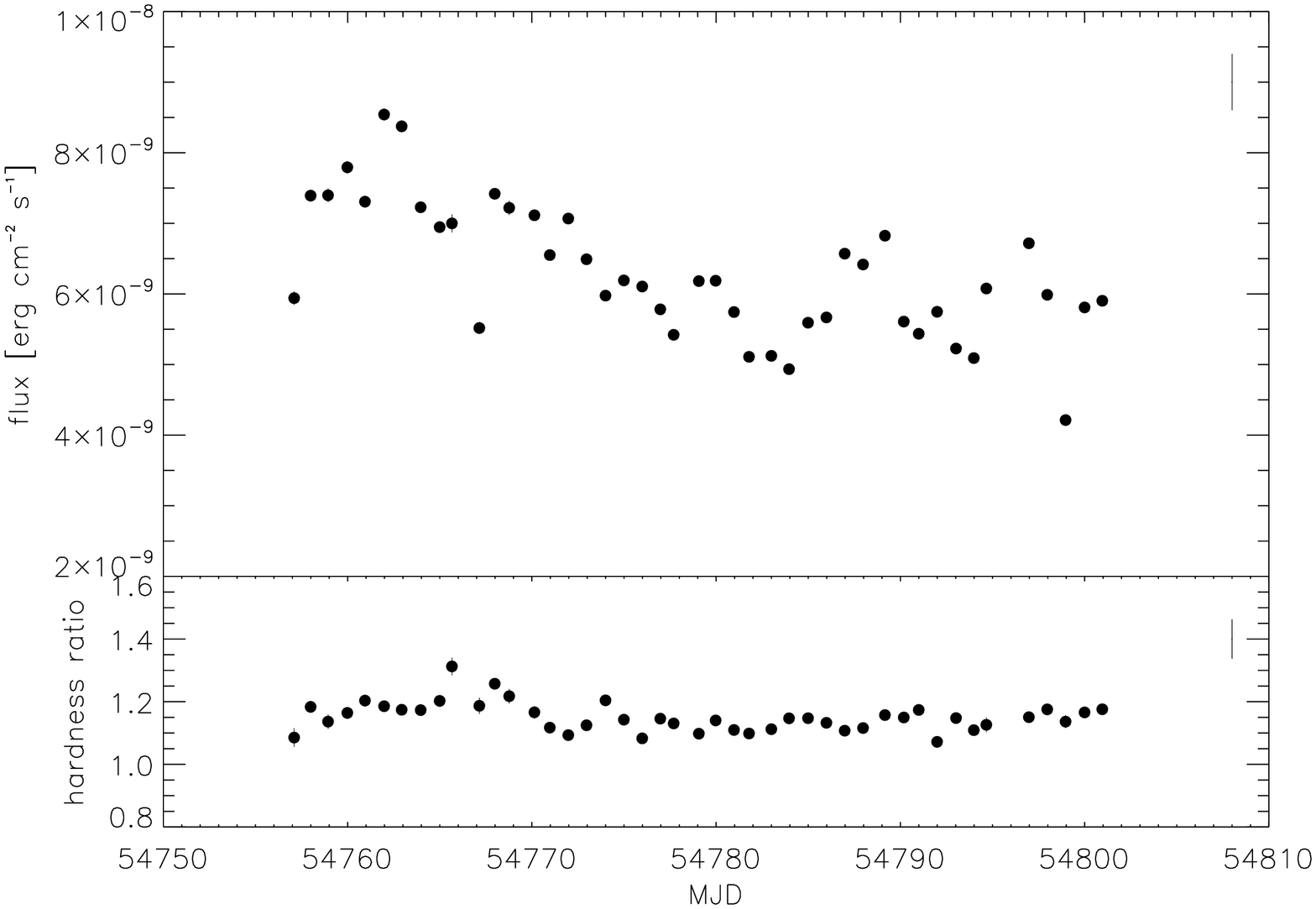}} \caption{SuperAGILE daily
lightcurve of Cyg X-1 during the observing blocks 1 (top left), 2
(top right), 3 (bottom left), and 4 (bottom right). Each point
represents the normalized rate estimated in one day (with exposure
of $\sim 40$ ks). The rate values are converted into units of
$\CGS$ considering the source average spectrum. The hardness ratio
is computed as the ratio of the emission in the HE band (25 -- 50
keV) to the LE one (20 -- 25 keV). Only the statistical
uncertainty is shown on the points in the plots. The average
systematic uncertainty is plotted in the top right corner of each
plot.} \label{fig:SuperAGILE_daily}
\end{figure*}

We studied the source variability in the SuperAGILE data with the
sampling time of one day, by accumulating the images separately in
two energy bands, low energy (LE) from 20 to 25 keV and
high-energy (HE) from 25 to 50 keV, with a net exposure of $\sim
40$ ks for each image. A hardness ratio is computed as the rate in
HE divided by the one in LE. The plots of the normalized rate and
hardness ratio are shown in Fig. \ref{fig:SuperAGILE_daily}. We
find that, during the observing blocks 1, 2 and 4, the hardness
ratio is roughly constant despite rate variations by roughly a
factor of $\sim 2$, with a few exceptions; for example, between
MJD 54441 and 54443, the hardness ratio increases in coincidence
with an enhancement of the counting rate after a local minimum,
owing to a flux reduction of about 50 \% on MJD 54442. In the
observing period 3, the hardness ratio shows a general linear
increase, with short-term fluctuations superimposed, corresponding
to a similar trend in the rate. The comparison with the data in 2
-- 12 keV (see Fig. \ref{fig:SA-XTE}) shows that the soft X-ray
flux decreases at the same time thus confirming the higher
hardness.

\begin{figure*}[h!]
\centering \subfigure{\includegraphics[angle=90,
width=.45\textwidth]{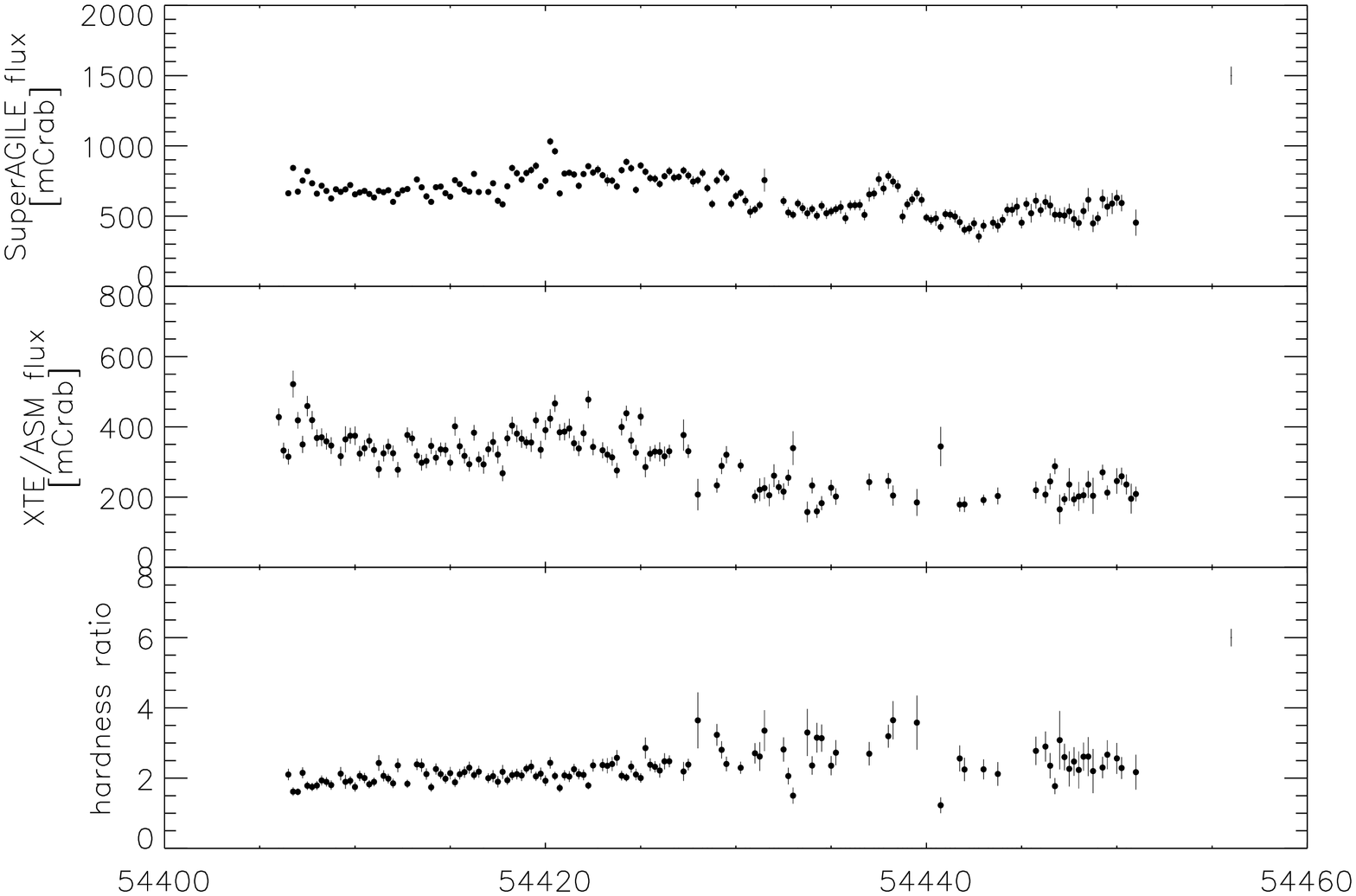}}
\subfigure{\includegraphics[angle=90,
width=.45\textwidth]{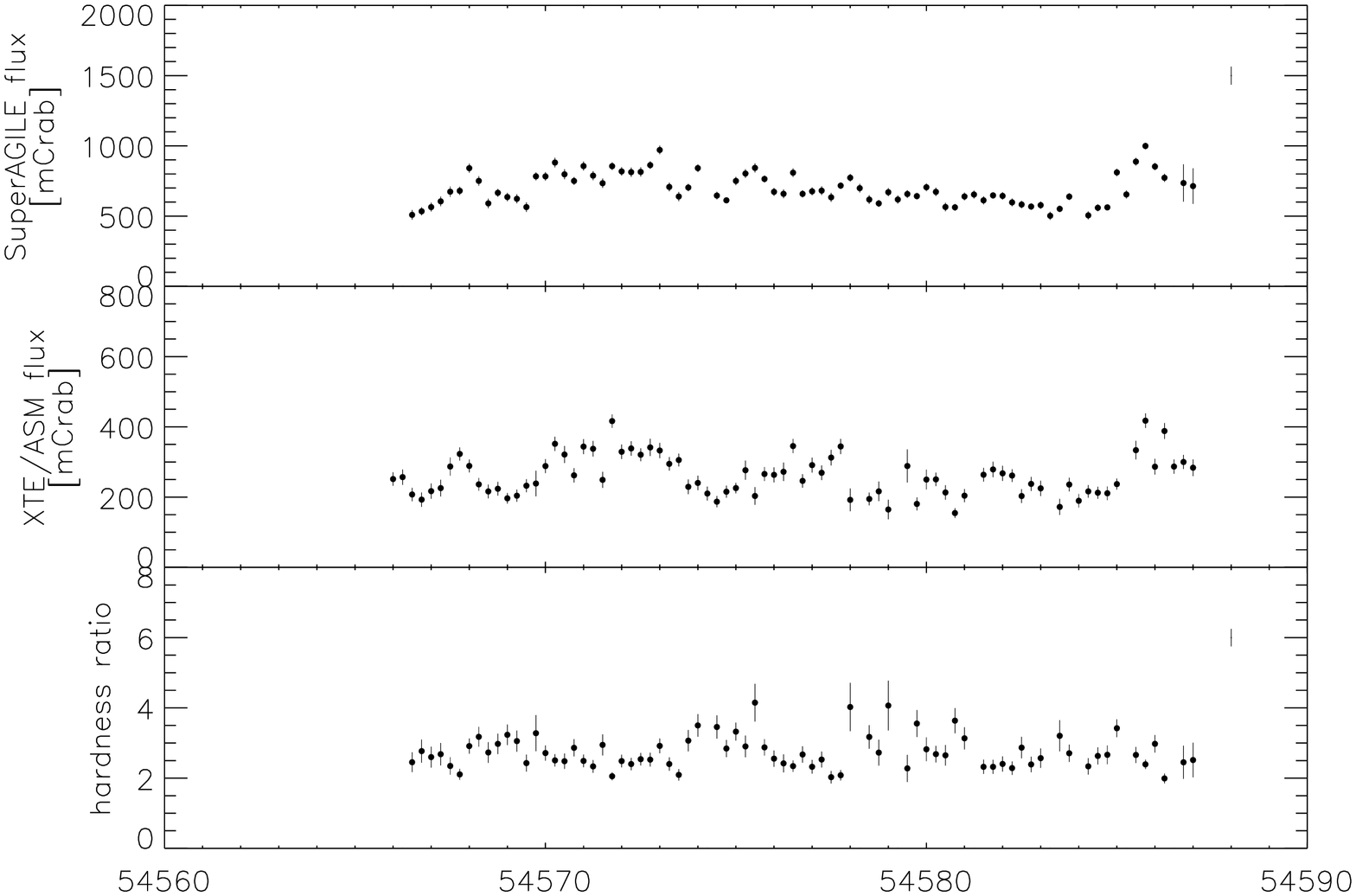}} \subfigure{
\includegraphics[angle=90, width=.45\textwidth]{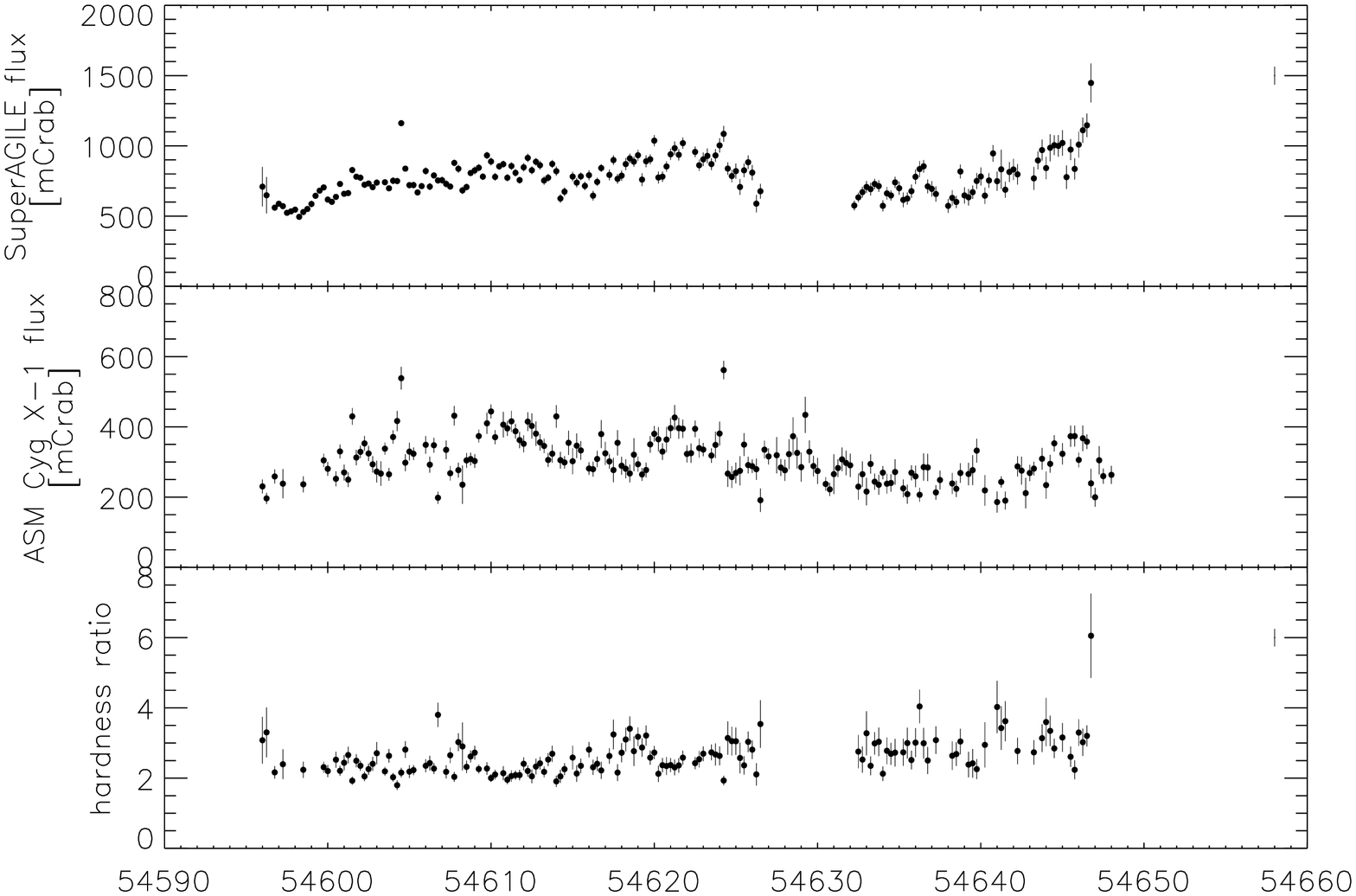}} \subfigure{
\includegraphics[angle=90, width=.45\textwidth]{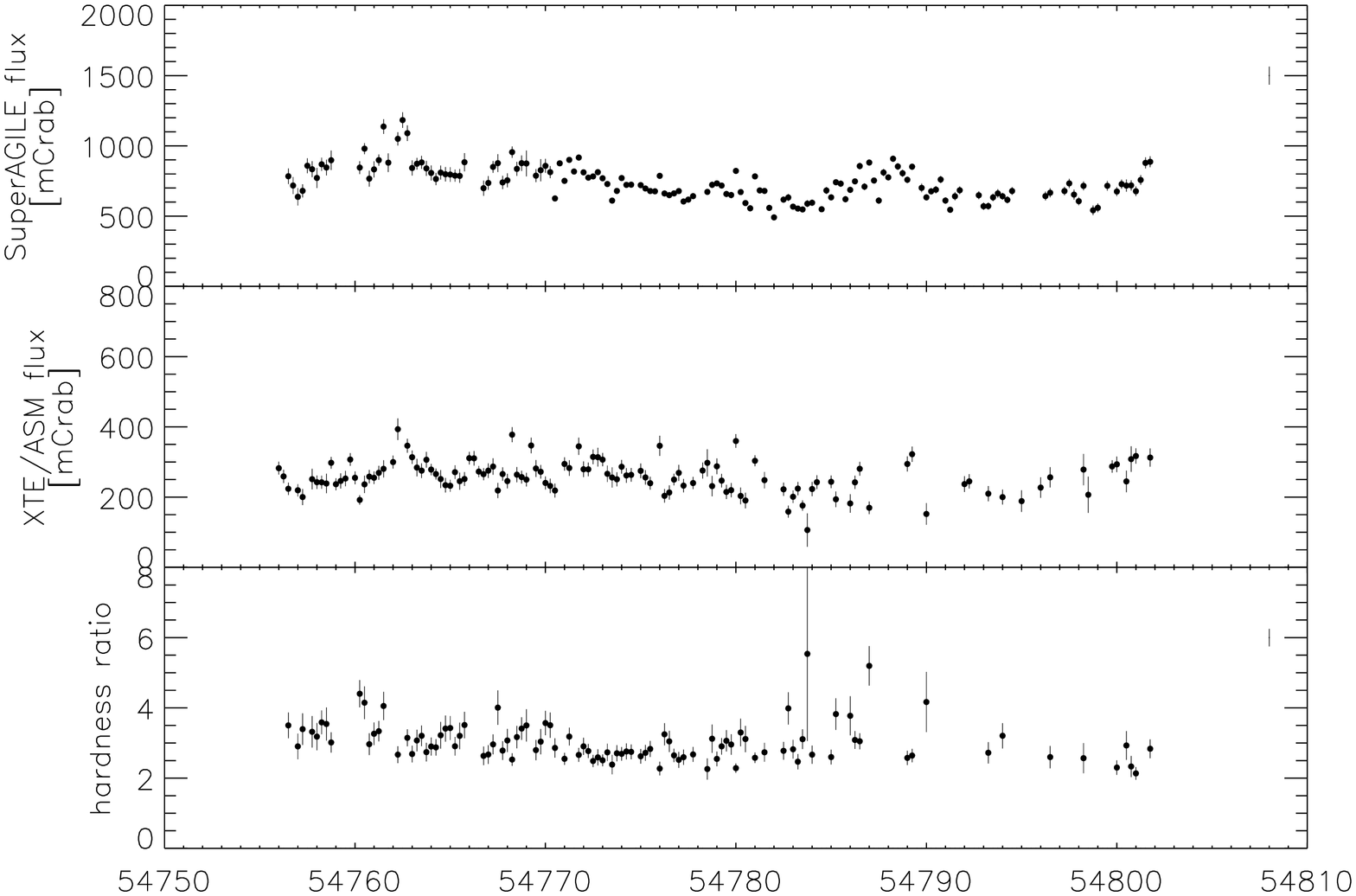}}
\caption{Superposition of the SuperAGILE (top panels) and RXTE/ASM
(middle panels) lightcurves of Cyg X-1 and hardness ratio (bottom
panels) during the observing periods 1 (top left), 2 (top right),
3 (bottom left), and 4 (bottom right). The arrows show the
position of periods of increase in the source emission that are
visible in the hard X-ray band but not in the soft one. Both
lightcurves in each period are accumulated on the same time
interval with a bin size of six hours. Only the statistical
uncertainty is shown on the points in the plots of the SuperAGILE
flux and of the hardness ratio. The average systematic uncertainty
is plotted in the top right corner of each plot.}
\label{fig:SuperAGILE-ASM_lc}
\end{figure*}

We investigated the contribution of the emission mechanism at
different energy ranges by studying the correlation between the
soft (from the RXTE/ASM public archive) and hard (SuperAGILE
observation) X-ray bands. We removed the dips simultaneous to the
inferior conjunction of the binary system \citep[observed by
][]{Mason_et_al_1974,Holt_et_al_1979,Priedhorsky_et_al_1995}, with
the stellar companion passing between the black hole and the
observer along the line of sight, using the orbital solutions for
HDE 226868 reported by \citet{La_Sala_et_al_1998}, with a period
of 5.5998 days. We then rebinned the lightcurves of both
instruments on a common timescale with the same bin size of six
hours, in order to increase the statistic of the measurement and
improve the chance of having at least one ASM measurememt in the
bin. From the superposition of the lightcurves and the hardness
ratio for each observing block, shown in Fig.
\ref{fig:SuperAGILE-ASM_lc}, we can see that there are intervals
of time of increased fluctuation in the hardness ratio. For
example in the OB1, the hardness ratio is almost constant between
MJD 54406 and 54427, and its variability increases until MJD
54442. Simultaneous to the increased variability, we can see that
the hard X-ray flux shows a flaring episode of about five days
duration reaching $\sim 800$ mCrab around MJD 54438, while the
soft X-ray flux has a decreasing trend. A similar behaviour is
shown in the OB4, with an increase in the hardness ratio between
MJD 54785 and 54793, corresponding to a bump in hard X-rays,
peaking around 500 mCrab on MJD 54788, while the soft X-ray
emission is almost constant at $\sim 250$ mCrab level. At the end
of the OB3, the flux measured by SuperAGILE increases up to $\sim
1500$ mCrab while the measure by RXTE/ASM remains between 200 and
400 mCrab, but in this period the source is at an off-axis angle
of about $32 \degree$, where the flux estimation is less accurate.
Unfortunately, the RXTE/ASM data are sparse in these time
intervals and consequently we cannot draw any serious conclusion.
To verify if we can find any variation in the source photon index,
we extracted the energy spectra of these episodes, but we did not
find any significant differences from the ordinary values. For
example, the photon index of the power-law that fits the spectrum
accumulated in the OB4 between MJD 54787 and 54789 is consistent
with the values reported in Table \ref{table:spectra}.

\begin{figure}
\centering
\includegraphics[angle=90, width=9. cm]{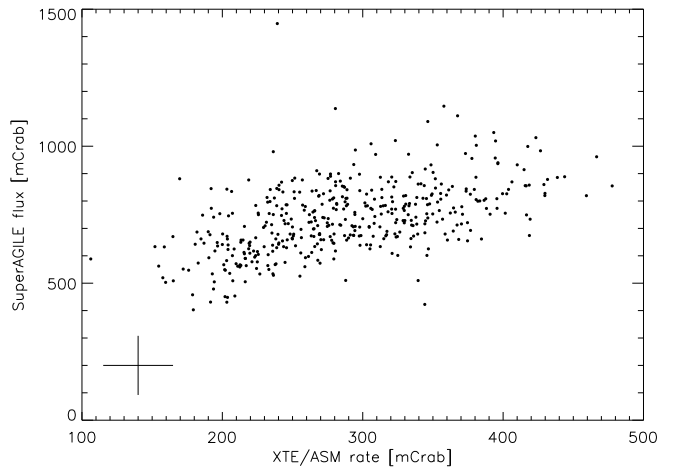}
\caption{Scatter plot of the SuperAGILE (20 -- 50 keV) vs RXTE/ASM
(2 -- 12 keV) flux, both separately converted into Crab units and
rebinned on a common timescale with a bin size of six hours. The
typical uncertainty is plotted for clarity only in the lower left
corner.} \label{fig:SA_vs_XTE_rebin}
\end{figure}

To investigate possible correlations between the source emission
in soft and hard X-rays, we show in Fig. \ref{fig:SA_vs_XTE_rebin}
the scatter plot of the fluxes of SuperAGILE and RXTE/ASM, each
one converted into Crab units for clarity. A certain degree of
correlation appears in the plot, although the value of the
correlation coefficient is 0.03 and does not allow establishment
of a formally significant correlation, even with more than 600
degrees of freedom. In the lower lefthand region of the plot
(between $\sim 150$ and $\sim 250$ mCrab values for the ASM rate),
we can see the data corresponding to the intervals in which the
source emission is increased in the hard X-ray range and is almost
constant in soft X-rays. We also computed the cross-correlation of
the RXTE/ASM and SuperAGILE lightcurves (both accumulated with six
hours resolution), and we did not find any time lag greater than
the bin size of six hours. The degree of correlation at zero time
lag is higher in the OB1 (0.73) and OB2 (0.69) than in the OB3
(0.54) and OB4 (0.55).

\subsection{Analysis of the structure function}

The SuperAGILE data are not well-suited to timing analysis of Cyg
X-1 using the standard tools, e.g. the fast fourier transform. In
fact, the measured rate is a few counts per second, depending on
the position of the source in the field of view, and the
occultation of the source by the Earth produces a discontinuity
(and consequently a windowing effect) on the time series.

As explained for example by \citet{Hughes_et_al_1992}, a useful
tool for studying the time variability of a physical quantity with
such sampling characteristics is the first-order structure
function, simply referred to as ``structure function'' or SF. This
function is commonly used in the analysis of time series
\citep[see e.g.][]{Rutman_1978}, has been introduced in astronomy
by \citet{Simonetti_et_al_1985} mainly for the study of active
galactic nuclei, and is also insensitive to temporal gaps in the
data. The insensitivity of the structure function to gaps in the
data is extremely important for satellites in low earth orbit,
such as AGILE, whose observations are interrupted by the
occultation of the source by the Earth and by the passage of the
satellite through the South Atlantic Anomaly.

The first-order SF is defined by \citet{Simonetti_et_al_1985} as
$D^1_f(\tau) \equiv \langle [f(t+\tau)-f(t)]^2 \rangle$ and in our
case it can be estimated as

\begin{equation}\label{eq:SF}
    D^1_f(\tau) = \frac{1}{N(\tau)} \sum [f(t+\tau)-f(t)]^2
\end{equation}

\noindent where the brackets $\langle \rangle$ indicate the
average, $\tau$ is the time lag, $f(t+\tau)$ and $f(t)$ represent
the flux at time $t+\tau$ and $t$, respectively, and $N(\tau)$ is
the number of pairs considered in the average.

A plateau at small time lags (often referred to as ``horizontal
branch'') is common of the SFs of the active galactic nuclei in
optical \citep[see for example][]{Paltani_et_al_1997} and gamma
rays \citep[see][]{Nandikotkur_et_al_1997} and is generally
produced by the white noise (uncorrelated flux) of the source
emission, yielding a constant value independent of the time lag. A
monotonic region is found at increasing lags, where the SF
approximately follows a power-law, $D^1_f(\tau) \propto \tau^{2H}$
\citep[see for example][]{Nandikotkur_et_al_1997}. The parameter
$H$ is also known as the Hurst exponent and is used in describing
non-stationary processes. Systems with $0.5 < H < 1$ are
\textit{persistent}: an increasing trend in the past is probably
followed by a similar increasing trend in the future. Conversely,
if $0 < H < 0.5$ the process shows \textit{antipersistence} and an
increase in the past is probably followed by a decrease in the
future. Finally, the minima in the structure function indicate
typical timescales of the system \citep{Nandikotkur_et_al_1997}.

\begin{figure}
\centering
\includegraphics[angle=90, width=9. cm]{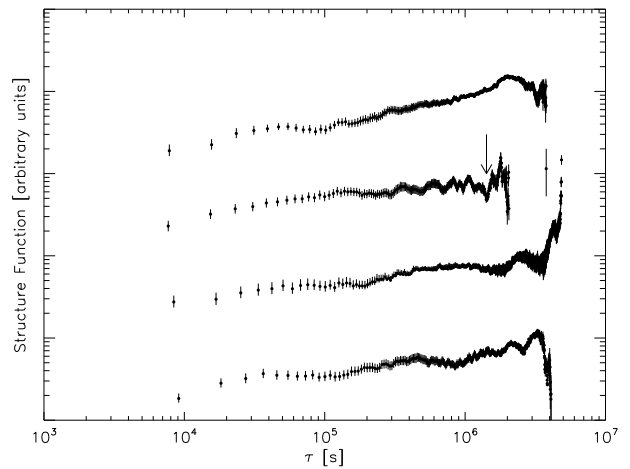}
\caption{Superposition of the first-order structure function of
the SuperAGILE flux extracted from the orbital integration. From
top OB1 to OB4 are shown. The arrow mark the position of the
minimum}.\label{fig:SF_orbital}
\end{figure}

\begin{figure}
\centering
\includegraphics[angle=90, width=9. cm]{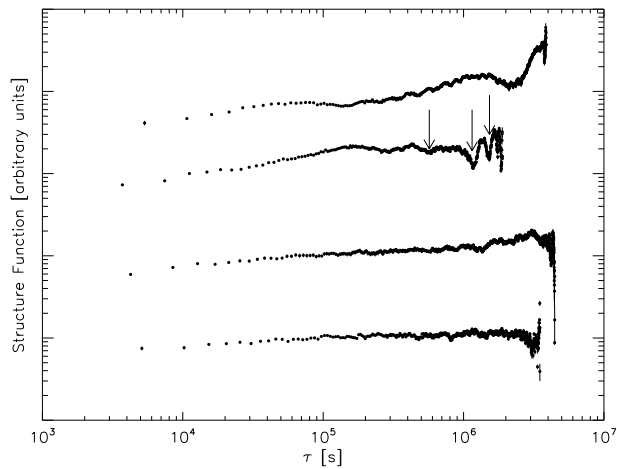}
\caption{Superposition of the first-order structure function of
the RXTE/ASM flux extracted from the ``dwells'' of 90 s duration.
From top OB1 to OB4 are shown. The uncertainty on the SF is
generally smaller than the symbol used in the plot. The arrows
mark the position of the minima.}\label{fig:SF_XTE}
\end{figure}

We computed the structure function of the normalized rate of Cyg
X-1, extracted separately in the four observing blocks and the two
energy bands (20 -- 50 keV of SuperAGILE and 2 -- 12 keV of
RXTE/ASM), estimating the rate with the time resolution of one
orbit ($\sim 3$ ks of net exposure) for SuperAGILE and one dwell
($\sim 90$ s) for RXTE/ASM. The SFs of the SuperAGILE data are
shown in Fig. \ref{fig:SF_orbital}, from top (OB1) to bottom
(OB4). Similarly, the structure function of RXTE/ASM data is shown
in Fig. \ref{fig:SF_XTE} with the same disposition of OB1 on top
and OB4 at the bottom.

The first evident feature in the structure function in both energy
bands is the absence of the ``horizontal branch'', indicating that
the variability of Cyg X-1 is correlated even at lags of the order
of $10^4$ s, the smallest timescale sampled in our observations.
The time lag values in the data span more than two orders of
magnitude, from $\sim 10^4$ s up to $\sim 4 \times 10^6$ s, with
the OB2 the only exception, limited to $\sim 2 \times 10^6$ s,
since this observing period is shorter than the others.

The structure functions in the hard X-ray band show a monotonic
increase between time lags of about $10^4$ up to $\sim 10^6$ s.
Above $\sim 10^6$ s the curve is less smooth, many fluctuations
appear and, in some cases, a few minima are found. Similar
behaviour is qualitatively found in the soft X-ray band, and also
in this case the monotonic region extends between $\sim 10^4$ s
and $\sim 2 \times 10^6$ s. In the OB1 and OB2 the structure
function in the two energy bands are comparable, while in the OB3
and OB4 the SuperAGILE data are different from the RXTE/ASM data,
which show a flatter power-law, with less pronounced fluctuations.

We define the minima as these values for which the structure
function is more than $3 \sigma$ far from the smoothed value. With
this method we find minima only in the OB2, at $1.42 \times 10^6$
s (corresponding to 16.5 days) in SuperAGILE and at $0.57 \times
10^6$ s (6.7 days), $1.15 \times 10^6$ s (13.3 days), and $1.53
\times 10^6$ s (17.7 days) in RXTE/ASM. Interestingly, these three
minima are all multiples of the same $\Delta t$ of 6.7 days, of
which 13.3 days is the double and 17.7 days has a ratio of 4/3.
The minimum in the SuperAGILE data has a ratio of 5/2 with the
$\Delta t$ of 6.7 days. We do not find minima in the other
observing periods, only fluctuations near the time corresponding
to half of the duration of the observation, where the scattering
of the data increases. This is the case, for example, of the
broad ``valleys'' in the SuperAGILE SFs of the OB3 (around gap of
$4 \times 10^6$ s) and OB4 (about $3 \times 10^6$ s) or the wide
fluctuation in the RXTE/ASM sample of the OB1 (between 2 and 3
$\times 10^6$ s).

To confirm our results in hard X-rays, we repeated the analysis of
the structure function using the publicly available Swift/BAT
data, in an energy band similar to SuperAGILE (15 -- 50 keV).
Searching with the same significance level, we found minima at
$0.42 \times 10^6$ s (4.9 days), $0.99 \times 10^6$ s (11.5 days),
and $1.42 \times 10^6$ s (16.5 days) in the OB2. The last ($1.42
\times 10^6$ s) is at the same position as the one found in the
SuperAGILE data. The analysis of the Swift/BAT data corresponding
to the OB1, OB3, and OB4 confirms the absence of minima in the SF.

To verify that the structure function is indeed insensitive to
windowing and aliasing of the time series, \citep[e.
g.][]{Hughes_et_al_1992}, we computed the SF of a simulated time
series. At the same time values of the OB2, shown in Fig.
\ref{fig:SuperAGILE_orbital}, we simulated the flux by extracting
with a random uniform distribution between the minimum and maximum
measured values. Then, we computed the structure function of this
simulated time series and applied the same algorithm used to find
the position of the minima. After $10^4$ simulations, our
algorithm found a fake minimum only in 71 occurrences, which
is the 0.7 \%, thus confirming the goodness of our results.

Finally, we fitted the function $\log[D^1_f(\tau)]$, independently
in the four observing blocks and the two energy bands, to estimate
the Hurst exponent $H$. Only in two cases, OB3 in the SuperAGILE
data and OB1 in the RXTE/ASM data, the structure function shows a
break and it is not possible to apply the same fit to the complete
curve, thus the interval has been split in two, fitted separately.
In the other cases, the structure function is described by a
single fit. The ranges of time lag in which the fits are applied
and the resulting value of the $H$ exponent are reported in Table
\ref{table:SA_SF} (SuperAGILE data) and Table \ref{table:XTE_SF}
(RXTE/ASM data). In all the observing blocks and in both energy
bands, $H$ is always significantly lower than the critical value
of 0.5, indicating \textit{antipersistence} in the source
emission.

\begin{table*}[h!]
\caption{Analysis of the structure function in the hard X-ray band}             
\label{table:SA_SF}      
\centering                          
\begin{tabular}{c c c c}        
\hline\hline                 
Obs.   & start lag & end lag & $H$  \\    
period & [s]       & [s]     &      \\    

\hline                        

OB1 & $10^{4.9}$ & $10^6$     & 0.185 \\
OB2 & $10^{3.9}$ & $10^{5.2}$ & 0.147 \\
OB3 & $10^{3.9}$ & $10^{4.9}$ & 0.115 \\
    & $10^{5.2}$ & $10^{5.6}$ & 0.210 \\
OB4 & $10^{4.9}$ & $10^{5.7}$ & 0.170 \\

\hline                                   
\end{tabular}

\caption{Analysis of the structure function in the soft X-ray band}             
\label{table:XTE_SF}      
\centering                          
\begin{tabular}{c c c c}        
\hline\hline                 
Obs.   & start lag & end lag & $H$  \\    
period & [s]       & [s]     &      \\    

\hline                        

OB1 & $10^{3.7}$ & $10^{4.7}$ & 0.144 \\
    & $10^{5.2}$ & $10^{6}  $ & 0.186 \\
OB2 & $10^{4}$   & $10^{5.2}$ & 0.148 \\
OB3 & $10^{4.2}$ & $10^{5.2}$ & 0.066 \\
OB4 & $10^{4.2}$ & $10^{5.2}$ & 0.050 \\

\hline                                   
\end{tabular}
\end{table*}

\subsection{Alternative methods of timing analysis}

We adopted additional and alternative methods of timing analysis
to verify the timescales of the system found in the structure
function: Lomb-Scargle periodogram and autocorrelation function.
These tools were chosen to comply with the structure of our time
series, which is very unevenly spaced.

The Lomb-Scargle periodogram is an algorithm specifically
developed to calculate the power spectrum of such non-uniformly
spaced data
\citep[][]{Lomb_1976,Scargle_1982,Horne_Baliunas_1986}. This tool
has been already adopted in the analysis of the long-term
variability of Cyg X-1, for example by \citet{Rico_2008} and
\citet{Benlloch_et_al_2004}, to search for the superorbital
period. Only some of the timescales from the structure function
are also present in the periodogram of the OB2: 6.7 and 13.3 days
in the RXTE/ASM data and 4.9 and 16.5 days in the Swift/BAT data.
Other timescales, in contrast, are not found in the Lomb-Scargle
diagram, such as the 16.5 days in SuperAGILE, the 17.7
days in RXTE/ASM and the 11.5 days in Swift/BAT.

By using the Lomb-Scargle periodogram, \citet{Rico_2008} finds a
superorbital modulation in the flux of Cyg X-1 with a period of
$326 \pm 2$ days. The SuperAGILE data span a total duration of
$\sim 395$ days, but the observing periods are shorter and not
evenly distributed during this time. For this reason we cannot
significantly test the presence of such superorbital modulation,
as confirmed by our attempts to fold our data on the periods
mentioned above.

We calculated the autocorrelation function from the SuperAGILE and
RXTE/ASM data in the four observing blocks (defined in Table
\ref{table:observation}) after removing the dips produced by the
inferior conjunction of the binary system and after rebinning the
data at six hours. Similar to \citet{Maccarone_et_al_2000}, we
computed the autocorrelation function in three energy bands: 20 --
50 keV of SuperAGILE, 1.5 -- 5 keV given by the sum of the A and B
bands of RXTE/ASM and 5 -- 12 keV from the ASM C band. Apart from
the peak at zero lag, no other peak is found, indicating that the
source emission does not show periodicity in all energy ranges.
Some differences in the width of the autocorrelation are found
depending on the energy band, but our analysis does not show any
definite trend.

\section{Results in the gamma ray energy band}

We analysed the GRID data to search for emission of Cyg X-1 above
100 MeV using different timescales: one day, five days, a single
observing block, and the complete AGILE observation. For the
timescales of one and five days, we analysed all the contiguous
time intervals included in the four observing blocks
independently. Compared to the real time, the live time is
typically $\sim 40$ ks and $\sim 200$ ks for the two types of
integrations, respectively, owing to the Earth occultation and the
satellite passages through the SAA. Cyg X-1 is not detected as a
point-like source on any timescale and the $2 \sigma$ upper limit
depends on the duration of the observation: $\sim 100 \times
10^{-8} \; \mathrm{ph \; cm^{-2} \; s^{-1}}$ in one day, and $\sim
50 \times 10^{-8} \; \mathrm{ph \; cm^{-2} \; s^{-1}}$ and $\sim
30 \times 10^{-8} \; \mathrm{ph \; cm^{-2} \; s^{-1}}$ in one
observing block. The value of the upper limit also depends on the
position of the source within the GRID field of view (see Table
\ref{table:observation}), which influences the effective area,
particularly in the last part of the OB1, where the upper limits
are $\sim 30$ \% higher.

\begin{figure}
\centering
\includegraphics[width=9. cm]{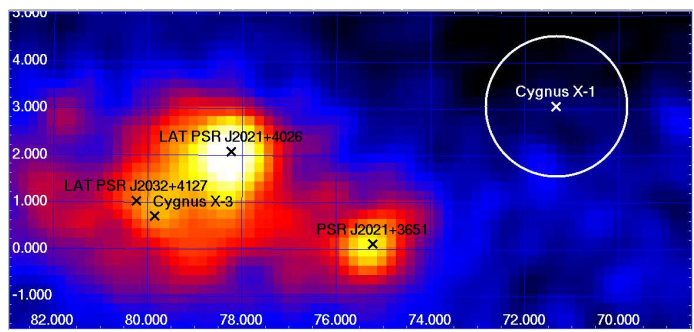}
\caption{Sum of the GRID count maps of the four observing periods,
The data are smoothed with a Gaussian shape of 3 pixel standard
deviation, and the circle at the position of Cyg X-1 has a radius
of $1.5 \degree$, corresponding to the instrument point spread
function at $\sim 200$ MeV.} \label{fig:GRID_cts_map}
\end{figure}

To find our deepest upper limit on the source flux, we accumulated
the maps of the whole observation, as shown in Fig.
\ref{fig:GRID_cts_map}. The circle at the position of Cyg X-1 has
a radius of $1.5 \degree$, corresponding to the GRID point spread
function at $\sim 200$ MeV. The source is not detected and the
upper limit above 100 MeV is $\mathrm{\sim 5 \times 10^{-8} \; ph
\; cm^{-2} \; s^{-1}}$. From the gamma ray map, it is possible to
see that the closest source is PSR J2021+3651, located at a
distance of $4.6 \degree$ from Cyg X-1, well outside the point
spread function. We can then exclude the contamination by field
sources.

\section{Discussion}

The AGILE mission observed Cyg X-1 $\sim4$ month-long continuous
periods of time during the science verification phase and the
observing cycle 1, between July 2007 and December 2008, for a
total net exposure to the source of about 6 Ms. Because of the
AGILE pointed strategy, during these observing periods the
monitoring of the source is continuous, interrupted only by the
periodic occultation by the Earth. Under this condition, the
SuperAGILE data on Cyg X-1 are a unique set in the hard X-rays.
Other hard X-ray experiments provided monitoring data on Cyg X-1,
but all of them are either (i) sparse, short snapshots of the
source, e.g. CGRO/BATSE \citep[][]{Crary_et_al_1996,
Ling_et_al_1997,Brocksopp_et_al_1999,Zdziarski_et_al_2002} and
Swift/BAT \citep[][]{Rico_2008} or (ii) short pointed observations
\citep[e.g., INTEGRAL/ISGRI: ][]{Cadolle_et_al_2006}. By taking as
a reference a colour-colour diagram built with the data of the
RXTE/ASM in the soft X-ray energy range, we verified that Cyg X-1
remained in its low-hard state the whole time.

\subsection{X-ray variability on short timescales}

We used the SuperAGILE data to systematically investigate the
source variability over our 160 days of monitoring on different
timescales, from $\sim 300$ s to the $\sim 1$-hour timescale, up
to week-long time segments. For the time periods when the source
exhibited a significant variability on the minute timescale, we
also (unsuccessfully) searched for potential flares on the
tens-of-second to subsecond-timescale in the SuperAGILE raw
lightcurves. Such variability has been observed in the past by
\citet{Gierlinski_Zdziarski_2003} on the ms timescale in the soft
X-rays with RXTE/PCA, and by Golenetskii et al. (2003) on the hour
timescale in the hard-X/gamma rays using the experiments of the
InterPlanetary Network. Both types of variability were observed
with the source in both soft and hard spectral states in a flux
range of a $few \times 10^{-7} \; \CGS$ (corresponding to a
luminosity of $\sim 10^{38} \; \mathrm{erg \; s^{-1}}$, assuming a
distance of 2 kpc) in the energy ranges 3 -- 30 keV and 15 -- 300
keV, respectively. During our observations we detected several
flares on timescales of a few hours to half a day, at times
corresponding to MJD 54420.4, 54585.6, 54604.6, 54607.9 and
54646.5. Contrary to the cases cited above, which involved flux
variability of one order of magnitude, these flares correspond to
flux increases of approximately a factor of 2 -- 3, passing from
$\sim 5-6 \times 10^{-9} \; \CGS$ to maxima of about $\sim 1.4
\times 10^{-8} \; \CGS$, as measured on the hour timescale in the
20 -- 50 keV energy range. Results similar to ours were also
reported by \citet{Brocksopp_et_al_1999} in a study based on data
of BATSE and RXTE/ASM, and by \citet{Malzac_et_al_2008} using
INTEGRAL/ISGRI (in the energy band 18 -- 40 keV).  In particular,
the 18 May 2008 (MJD 54604) flaring episode reported by
\citet{ATel_1533}, with a duration of about 8 hours, is among the
flares observed by SuperAGILE during the OB3. The flux measured by
SuperAGILE in 20 -- 50 keV between 15:02 and 15:36 UT, when
INTEGRAL/ISGRI detected the maximum of the source emission, is
$\sim 1.5$ Crab.

\subsection{Long-term variability}

Interestingly, with the exception of the flares discussed in the
previous section, during the SuperAGILE monitoring the hard X-ray
flux of Cyg X-1 remained quite stable around $\sim 5-6 \times
10^{-9} \; \CGS$, with fluctuations on much longer timescales. We
investigated this long-term variability by using different
analysis tools. The SuperAGILE data are characterized by long gaps
between the observing periods, and shorter gaps due to the Earth's
occultation in every orbit. For this reason we applied the
analysis of the structure function for the first time in the study
of a Galactic binary source, but frequently used for the active
galactic nuclei in the radio and gamma ray energy bands \citep[see
for
example][]{Hughes_et_al_1992,Simonetti_et_al_1985,Nandikotkur_et_al_1997}.
We performed a comparative analysis in the hard and soft X-rays,
complementing the SuperAGILE data with those of RXTE/ASM available
in the public archive. Since the slope of the structure function
is connected to the probability of long duration trends in the
emission mechanism, while the minima in the same function are
related to the typical timescales of the same process, we address
the two topics separately.

By using a logarithmic fit to the structure function versus the
time lag, we find that, throughout the whole period of our
analysis, the Hurst exponent $H$ is significantly below the
critical value of 0.5, indicating that the source is
\textit{antipersistent}. This \textit{antipersistence} is a
property of random non-stationary processes and indicates that an
increase in the past is likely to be followed by a decrease in the
future \citep{Karner_2005}. \textit{Antipersistent} systems are
not common in nature and show a dominant negative feedback at work
that produces ``oscillations'', i. e. repetition of similar
features but without periodicity, because the distance in time is
not constant \citep[see for example][]{Koutsoyiannis_2008}.
\citet{Gil-Alana_2005} and \citet{Karner_2005}, for example, find
\textit{antipersistence} in the climatological study of the time
series of the temperature in various layers of the Earth
atmosphere and discuss possible mechanisms of negative feedback
(ice albedo, water vapour and clouds). On the other hand,
\citet{Alvarez-Ramirez_et_al_2002}, always by means of the Hurst
exponent, find that the crude oil market is a \textit{persistent}
process with long-run memory effects and different timescales at
work: the ``oscillations'' from days to weeks are apparently
generated by the action of market speculators and are superimposed
on a long-time trend, with characteristic timescale of weeks to
three months.

The description of the \textit{antipersistent} systems matches the
behaviour of Cyg X-1, whose luminosity does not seem to vary on
long timescales and shows non-periodic ``oscillations'' around an
average value of $\sim 2 \times 10^{37} \; \mathrm{erg \; s^
{-1}}$. This aperiodic variation is well known especially in the
low/hard spectral states, \citep[see for example][and references
therein]{Makishima_et_al_2008}. A complete discussion of the
possible negative feedback mechanisms in the emission of Cyg X-1
would require a detailed analysis of the interactions between the
various components at work (stellar wind, accretion disk,
Comptonizing plasma, and relativistic jet) and is beyond the scope
of the present paper.

The position of the minima in the structure function is an
indication of the typical timescales of the source
\citep{Nandikotkur_et_al_1997}. In our analysis we only found
minima in the structure function of the observing block 2 (April
2008), but at different positions in SuperAGILE (16.5 days) and
RXTE/ASM (6.7, 13.3, and 17.7 days). The analysis of the structure
function of the Swift/BAT data in the same periods confirms the
SuperAGILE minimum at 16.5 days and is accompanied by two more
minima on slightly different timescales (4.9 and 11.5 days). The
minima that we detected do not correspond to any known timescale
of Cyg X-1, which has an orbital period of $\sim 5.6$ days
\citep{La_Sala_et_al_1998}, and are not confirmed by other
techniques of timing analysis. In fact, we do not detect any
periodicity using the autocorrelation function, and we found only
some of the timescales in the Lomb-Scargle periodogram. An
inspection of the lightcurves of the OB2 reveals that the
timescales found in the structure function correspond to features
in the time series: 6.7 days, which is also present in the
Lomb-Scargle periodogram, is the average distance between the
minima in the soft X-ray lightcurve and 16.5 days is the
separation of the two maxima of the emission in hard X-rays. From
all these results we conclude that the timescales that we found in
the structure function correspond to emission features of the
specific observation period and not to general properties of the
source.

\subsection{The search for gamma ray emission}

Thanks to the long and continuous monitoring of AGILE, we were
able to investigate the possible emission of gamma rays from Cyg
X-1 on both short and long timescales. An investigation on the
daily timescale did not provide any significant detection. We then
increased the exposure time to five days, one observing block
($\sim 1$ month) and finally the whole 160-days observation. We
did not find any significant emission of gamma rays on any of
these timescales, the typical $2 \sigma$ upper limit being
$\mathrm{\sim 100 \times 10^{-8} \; ph \; cm^{-2} \; s^{-1}}$ on
the one-day and $\mathrm{\sim 50 \times 10^{-8} \; ph \; cm^{-2}
\; s^{-1}}$ on the five-day timescale. The tightest $2 \sigma$
upper limit derives from the integration over the whole
observation (6.1 Ms net exposure) and corresponds to $\mathrm{\sim
5 \times 10^{-8} \; ph \; cm^{-2} \; s^{-1}}$.

Recently, an episodic flare from Cyg X-1 was detected in gamma
rays by AGILE in an interval of time subsequent to that considered
in the present paper, and reported by \citet{Sabatini_et_al_2010}.
Between 15 and 16 October 2009 (MJD 55119.97 -- 55120.96), a flux
of $(232 \pm 66) \times 10^{-8} \; \mathrm{ph \; cm^{-2} \;
s^{-1}}$ above 100 MeV was measured, at a significance level of
$5.3 \sigma$ pre-trial and $4 \sigma$ post-trial, when the source
was in a hard spectral state. \citet{Sabatini_et_al_2010} also
extended the search for persistent emission over all the AGILE
archival data of the Cygnus Field, including the interval covered
by the present paper, and did not find any significant detection.
Given the longer integration time, the authors also obtained an
upper limit slightly deeper than found in our analysis, $\sim
3 \times 10^{-8} \; \mathrm{ph \; cm^{-2} \; s^{-1}}$, with an
integration time of about 300 days, spread over 2.5 years.

Cyg X-1 has also been detected at TeV energies by MAGIC
\citep{Albert_et_al_2007} during an $\sim$hour flare, at the
rising edge of a flaring episode detected in both soft and hard
X-rays. This rare, and so far unique, event occurred with the
source in a hard spectral state, as in our observations, although
reaching a flux of $\sim$2 Crab in hard X-rays, while our X-ray
data never show the source above 1.5 Crab. The MAGIC observation
shows a power-law spectrum between 150 GeV and 2 TeV with photon
index $3.2 \pm 0.6$, yielding a flux of $\sim 7 \times 10^{-11} \;
\mathrm{ph \; cm^{-2} \; s^{-1}}$. If we assume that the spectrum
can be extrapolated to the energy range 0.1 -- 1 GeV, as when the
emission site is far from the disk and the plasma, so that the
cross-section of the photon conversion into electron positron
pairs is negligible \citep[see][]{Zdziarski_et_al_2009}, we obtain
an expected flux value of $\sim 63000 \times 10^{-8} \; \mathrm{ph
\; cm^{-2} \; s^{-1}}$. Even ``diluting'' such a flux on a 24-hour
timescale, it would correspond to an expected day-averaged flux of
$\sim 3000 \times 10^{-8} \; \mathrm{ph \; cm^{-2} \; s^{-1}}$,
well above the GRID daily sensitivity of $\sim 100 \times 10^{-8}
\; \mathrm{ph \; cm^{-2} \; s^{-1}}$. Thus, regardless of the
spectral state of the source, our analysis can exclude the
presence of gamma ray (GeV-)flares similar to what is observed by
MAGIC at TeV energies, during the periods of the AGILE
observation.

Cyg X-1 was observed by the instruments aboard the \textit{Compton
Gamma Ray Observatory} on several occasions during the first three
cycles (1991 -- 1994), when the source was in a low/hard state
\citep{McConnell_et_al_2000}. COMPTEL measured an energy spectrum
described by a simple power-law with photon index $3.3 \pm 0.4$,
between 0.75 and 5 MeV. Above 5 MeV, the statistical quality of
the data does not allow placing more than an upper limit
consistent with the extrapolation of the spectrum at lower
energies. During the same observations, EGRET did not detect the
source, with an upper limit on the flux at $10 \times 10^{-8} \;
\mathrm{ph \; cm^{-2} \; s^{-1}}$ (100 -- 200 MeV). The deepest
AGILE upper limit \citep{Sabatini_et_al_2010} is about a factor of
three lower than EGRET. However, an extrapolation of the COMPTEL
spectrum above 100 MeV would provide an expected flux of $\sim 5
\times 10^{-9} \; \mathrm{ph \; cm^{-2} \; s^{-1}}$, quite
consistent with both of them. Our data, then, do not provide
evidence of any spectral cut-off above the COMPTEL energy
range.

Interestingly, Cyg X-1 was also detected by COMPTEL in high/soft
state, up to about 10 MeV energy \citep{McConnell_et_al_2002}. At
that time the EGRET instrument was switched off and no GeV
observation is available. The COMPTEL spectrum of the soft state
can be modelled with a single power-law of photon index 2.6
extending from 1 to 10 MeV. Extrapolating this energy spectrum to
the AGILE energy band would provide a flux expectation of $\sim30
\times 10^{-8} \; \mathrm{ph \; cm^{-2} \; s^{-1}}$, easily
detectable by the AGILE/GRID on a timescale of a few days. This
can be seen as an additional indication that Cyg X-1 most likely
remained in its low/hard spectral state during all the time of the
AGILE.

The AGILE upper limits on GeV-emission from Cyg X-1 range from
100$\times 10^{-8} \; \mathrm{ph \; cm^{-2} \; s^{-1}}$ (one day
timescale) to 3$\times 10^{-8} \; \mathrm{ph \; cm^{-2} \;
s^{-1}}$ ($\sim 2.5$ years average). Assuming a source distance of
2 kpc, \citep{Ziolkowski_2005} they correspond to an isotropic
source luminosity of $\sim 8 \times 10^{34}$ erg s$^{-1}$ and
$\sim 2 \times 10^{33}$ erg s$^{-1}$, respectively. Different
authors elaborated models for the high-energy emission from Cyg
X-1. \citet{Zdziarski_et_al_2009} proposed a model to explain the
short TeV flare detected in 2006 by MAGIC
\citep{Albert_et_al_2007}. They propose that high-energy photons
are generated by electrons accelerated to TeV energies close to
the central X-ray source. The interaction of such TeV photons with
the photon field of the X-ray source yields produces
$\mathrm{e^{\pm}}$ pairs. They find that, for initial photon
energies above $\sim$3 TeV, photons can travel far enough,
initiating extended pair cascades, in turn producing an observable
flux. The predicted photon energy spectrum depends on the energy
of the primary electron injection and on the stellar temperature.
Under the conditions observed in X-rays during the TeV flare
\citep[when the hard X-ray flux was only $\sim$two times higher
than what we observed with SuperAGILE, ][]{Malzac_et_al_2008}, the
model predicts a flux in the AGILE energy range as high as
$\sim4\times 10^{34}$ erg s$^{-1}$. This is a luminosity value
higher than the AGILE time-averaged upper limit on the persistent
emission, while it is still compatible (2 times less) with the
1-day upper limit.

Alternative models have also been proposed.
\citet{Araudo_et_al_2009} studied the interaction of the jet of
Cyg X-1 with the wind of the companion, expected to be clumpy, and
predicted the possible spectral energy distributions under
different conditions of the system. Since the radiation is
considered to be produced by particles accelerated in the shock
due to the interaction between the jet and a cloud, the emission
is predicted to be transient, with the duration timescale expected
in the range between minutes to a few hours. The frequency of
occurrence of such transient emission depends on the clumps'
number and size, up to appearing as a modulated steady emission.
Interestingly, under some magnetic field / clump size conditions,
the model predicts the radiative output due to inverse Compton
effect to reach luminosities in the AGILE energy range as high as
$\sim few \times 10^{34}$ erg s$^{-1}$, comparable to the AGILE
daily upper limits and nearly one order of magnitude higher than
the yearly-average limit. Also \citet{Orellana_et_al_2007} suggest
that the GeV -- TeV emission in microquasars may originate from
the interaction between the jet and the stellar wind of the
companion. In this case the inelastic interactions between the
relativistic protons in the jet and the cold protons of the
stellar wind produce charged ($\pi ^\pm$) and neutral ($\pi ^0$)
pions. The jet is assumed to be continuous and to contain a
randomly oriented magnetic field. Two main channels of gamma ray
production emerge from the model: the decay of neutral pions and
the emission by charged leptons, resulting from the decay of
charged pions and from the photon-photon interactions, producing
photons via synchrotron emission and inverse Compton scattering on
the low energy stellar photons. Similar to
\citet{Araudo_et_al_2009}, \citet{Orellana_et_al_2007} present the
spectral energy distribution (SED) expected in their model for the
two emission channels ($\pi ^\pm$ and $\pi ^0$) using the
parameters of Cyg X-1. The gamma ray luminosity produced by the
decay of the neutral pions depends on the subtended solid angle
and on the inclination of the jet toward the line of sight. Taking
an inclination of $30 \degree$ into account \citep[measured by][in
radio]{Gallo_et_al_2005} and a semi-opening angle of 0.1 radians,
the expected luminosity is more than three orders of magnitude
lower than the AGILE upper limit, at an average energy of 100 MeV.
On the other hand, the leptons emitted in the decay of the charged
pions can produce gamma rays via inverse Compton scattering of low
energy photons from the companion star. The interactions of gamma
rays with low energy photons may in turn produce relativistic
$e^\pm$ pairs that can Compton upscatter the low energy photons or
trigger an electromagnetic cascade. In the SED of such secondary
emission from charged pions, adopting two values of the magnetic
field, $10^4$ G and $10^5$ G, the luminosity at 100 MeV is
slightly above the $\sim 2 \times 10^{33} \; \mathrm{erg \;
s^{-1}}$ AGILE upper limit for the highest magnetic field ($10^5$
G) while the model is still compatible with our data for a lower
magnetic field, $10^4$ G, whose luminosity is about one order of
magnitude less than our upper limit.

\section{Summary and conclusions}

We reported the first campaign of observation of the hard spectral
state of Cyg X-1 in gamma rays ($>100$ MeV) on various timescales,
with exposures ranging from one day up to $\sim 160$ days. We
monitored the source simultaneously in hard (20 -- 50 keV) and
soft (2 -- 12) X-rays using SuperAGILE and RXTE/ASM. The
observation in hard X-rays by SuperAGILE shows the well-known
erratic variability of the flux of Cyg X-1. The analysis of both
the hardness ratio, estimated from the SuperAGILE flux in the two
energy bands 20 -- 25 keV and 25 -- 50 keV, and of the
colour-colour diagram, obtained from the public data of RXTE/ASM
following the method reported by \citet{Reig_et_al_2002}, does not
show any transition of spectral state. We adopted exposure times
ranging from $\sim 300$ s up to one day and found the typical
short time flickering of Cyg X-1 with intensity variations by
about a factor of two.

From the study of the first-order structure function, we did not
find any short lag plateau (``horizontal branch''). A power-law
behaviour is found, with the Hurst exponent smaller than 0.5,
denoting that the emission mechanism of Cyg X-1 is
\textit{antipersistent}, i. e. dominated by negative feedback. We
also found timescales from the minima in the structure function
but, from the analysis of the autocorrelation function and the
Lomb-Scargle periodogram, we derive that these timescales are more
probably related to particular patterns in the specific
lightcurve, such as the distance between the repeated minima or
the peaks of the emission, rather than to intrinsic properties of
the source.

Cyg X-1 is not detected as a point like source above 100 MeV, and
we find values of the upper limit ranging from $\sim 100 \times
10^{-8} \; \mathrm{ph \; cm^{-2} \; s^{-1}}$ in one day down to
$\mathrm{\sim 5 \times 10^{-8} \; ph \; cm^{-2} \; s^{-1}}$ for
the whole observation, about a factor of two lower than from the
EGRET data. We compared the luminosity derived from the AGILE
upper limit above 100 MeV (assuming a distance of 2 kpc to the
source) with various models of GeV -- TeV emission. The
predictions of the pairs' cascade model \citep[proposed by][to
explain the TeV flaring emission detected by
MAGIC]{Zdziarski_et_al_2009} are compatible with our upper limit
only for flaring emission (e. g. one-day timescale), while they
are about one order of magnitude higher when compared to the
year-long upper limit. We compared our results with the
predictions of two alternative models, involving hadronic
interactions between the cold matter of the stellar wind and the
relativistic jet, proposed by \citet{Araudo_et_al_2009} and
\citet{Orellana_et_al_2007}. The first model is more suitable for
flaring emission, while the second one can explain the persistent
emission better. We find that the luminosity in the $\sim 100$ MeV
range is compatible with the results of the AGILE monitoring for
either short transient or for selected model parameters.

\begin{acknowledgements}

The authors are grateful to L. Stella, G. Israel, and G. Matt for
useful suggestions and discussions. We also thank the anonymous
referee for stimulating us to significantly improve the quality of
the paper and the language editor for useful suggestions and
corrections. In the paper we used the Swift/BAT transient monitor
results provided by the Swift/BAT team and the RXTE/ASM public
data archive. AGILE is a mission of the Italian Space Agency, with
co-participation of INAF (Istituto Nazionale di Astrofisica) and
INFN (Istituto Nazionale di Fisica Nucleare). This work was
partially supported by ASI grants I/R/045/04, I/089/06/0,
I/011/07/0 and by the Italian Ministry of University and Research
(PRIN 2005025417). INAF personnel at ASDC are under ASI contract
I/024/05/1.

\end{acknowledgements}

\bibliographystyle{aa} 
\bibliography{13104}
\end{document}